\documentclass[preprint,aps]{revtex4}
\usepackage{graphicx}
\usepackage{amssymb}
\usepackage{slashed}
\newcommand{\be}{\begin{eqnarray}}
\newcommand{\ee}{\end{eqnarray}}
\usepackage[colorlinks]{hyperref}
\usepackage{cancel}

\usepackage{amsmath,amssymb}
\usepackage{mathtools}
\usepackage{float}
\usepackage{color}
\begin{document}

\title{Probing Transverse Momentum Dependent Parton Distributions in Charmonium and
Bottomonium Production}

\author{\bf Asmita Mukherjee and Sangem Rajesh}

\affiliation{ Department of Physics,
Indian Institute of Technology Bombay, Mumbai-400076,
India.}
\date{\today}

\begin{abstract}
\noindent
We propose the study of  unpolarized transverse momentum dependent gluon
parton distributions as well as the effect of linearly polarized gluons on transverse 
momentum and rapidity distributions of $J/\psi$ and $\Upsilon$ production within the 
framework of transverse momentum dependent factorization 
employing color evaporation model (CEM) in unpolarized proton-proton collision. We estimate the 
transverse momentum and rapidity distributions of $J/\psi$ and $\Upsilon$ at LHCb, RHIC  
and AFTER  energies using TMD evolution formalism.
\end{abstract}

\maketitle

\section{Introduction}\label{sec1}

In recent years, transverse momentum dependent parton distributions (TMDs)
and fragmentation functions (TMFs) have gained a lot of interest both
theoretically and experimentally. These are objects that play an essential
role for example in single spin asymmetries, where one of the colliding
beams/target is polarized. In order to gain information on TMDs, one needs
a process like semi-inclusive deep inelastic scattering (SIDIS) or Drell-Yan
(DY), where one observes the transverse momentum of a produced particle.
There are interesting theoretical issues associated with the TMDs like
universality and factorization \cite{tmde2}, and a lot of theoretical work
have been done in the past few years to shed light on these issues. However,
more experimental data are required to fully understand them. There are
interesting and useful data
from HERMES, COMPASS, JLab, as well as from Tevatron, LHC, Belle and BaBar
collaborations.  The ultimate goal is to obtain a global fit for the TMDs
using all data. However, the problem lies in the fact that the data from
SIDIS are at a low energy compared to the data for DY process \cite{lowPT}.
It has been shown that in the range $\Lambda_{QCD} << k_\perp << Q$ where
$k_\perp$  is the transverse momentum and $Q$ is the large momentum scale of
the process, radiative gluon emission play an essential role and these need
to be resummed, giving rise to evolution of TMDs \cite{tmde3}, whereas for
low $k_\perp$, non-perturbative physics dominates. To get a complete
picture, one would need to use a lot of data from different experiments in
different kinematics. Here it is important to first investigate the
unpolarized TMDs. Not only this is important to understand the behaviour of
TMDs in general over a large range of momentum scales, but also it is
important for spin-dependent studies as these lie in the denominator of spin
asymmetries. Information on quark and antiquark TMDs  can be obtained for
example from SIDIS, however, gluon TMDs can be best studied in $pp$ and
$p{\bar p}$ collisions.                 

It has been pointed out \cite{prd57} that gluons can be linearly polarized even inside an 
unpolarized hadron provided that gluons should have non-zero transverse momentum with respect to the
parent hadron. The gluon-gluon correlation function contains the information regarding linearly 
polarized gluons. Formally, gluon correlator of the unpolarized spin$-\frac12$ hadron is parameterized
in terms of leading twist transverse momentum dependent (TMD)  distribution functions \cite{prd63}
i.e., $f^g_1(x,{\bf k}_{\perp}^2)$ and $h^{\perp g}_1(x,{\bf k}_{\perp}^2)$. Here, $f^g_1$ represents
the likelihood of finding an unpolarized gluon with longitudinal momentum fraction $x$ and transverse 
momentum $k_{\perp}$ inside an unpolarized hadron and $h^{\perp g}_1$, Boer-Mulders function,
describes the distribution of linearly polarized gluons within the unpolarized hadron
(time-reversal even, or T-even).
In general, quarks can also be transversely polarized within the unpolarized hadron. 
Quark distribution function $h^{\perp q}_1$ is time-reversal odd (T-odd) function. Experimentally, large $\cos2\phi$ 
azimuthal asymmetry was observed in DY process \cite{N01,N02}. 
It was also noticed in SIDIS by EMC
\cite{EMC} and ZEUS \cite{ZEUS} experiments. As suggested in \cite{prd60}, the observed
$\cos2\phi$ azimuthal asymmetry can be explained by Boer-Mulders effect. Quark Boer-Mulders 
function ($h^{\perp q}_1$) has been explored in SIDIS process \cite{prd81} assuming a 
relation with Sivers function $f^{\perp q}_{1T}$ \cite{sivers1,sivers2}. It
was noticed in \cite{prd81}
that Cahn effect \cite{pl78,prd40} also leads to the $\cos2\phi$ asymmetry with a comparable 
contribution as that of Boer-Mulders effect. Moreover, antiquark Boer-Mulders function was estimated 
in DY process \cite{prd82} by measuring azimuthal asymmetry. Recent analysis on $h^{\perp q}_1$ has been
given in Ref. \cite{prd91}.

However, no experimental investigation to extract the gluon Boer-Mulders function ($h^{\perp g}_1$) 
has been carried out until date. Hence, the quantification of $h^{\perp g}_1$ is still enigmatic except
knowing its theoretical upper bound \cite{prl106}. Towards this, numerous 
proposals have been introduced to investigate $h^{\perp g}_1$, theoretically. The diphoton
production has been suggested to probe the linearly polarized gluons in
$pp\rightarrow\gamma\gamma +X$ at relativistic heavy ion collision (RHIC) \cite{prl107}.
Additionally, SIDIS and hadronic collision \cite{jhep} processes were also proposed to
probe $h^{\perp g}_1$ in heavy quark and dijet production respectively. It has been shown that
$h^{\perp g}_1$ causes the imbalance of dijet in 
hadronic collision \cite{prd80} through which an estimation of the average intrinsic transverse
momentum of the partons can be obtained. It was observed that Higgs boson transverse momentum 
distribution at LHC has been modified by the presence of linearly polarized gluons in the
unpolarized hadrons \cite{prd84,prl108,prl111}. The modified transverse momentum spectrum 
provides a way to determine whether the Higgs is a  pseudo-scalar or scalar  \cite{prl108}. 
The effect of linearly polarized gluons on the transverse momentum distribution in heavy quark
pair production ($\eta_{c,b}$, $\chi_{c0,b0}$ and $\chi_{c2,b2}$) has been studied 
in $pp$ collision \cite{prd86} using non-relativistic QCD (NRQCD).

TMD factorization framework for SIDIS and DY processes has now been derived \cite{tmde2}.
TMD functions depend on the intrinsic transverse
motion $k_{\perp}$  along with the longitudinal momentum fraction $x$ of the
partons whereas in collinear 
factorization the parton distribution functions (pdfs) depend only on $x$ in which the 
transverse momentum is integrated out. In addition, these depend on the
momentum scale $Q$. The transverse momentum ($P_T$) spectrum of a particular final state hadron in the scattering process
 seems to have Gaussian distribution \cite{lowPT}. This prompted the
assumption that  TMD pdfs exhibit
 Gaussian shape. A simple Gaussian model in which the TMD pdf factorizes into exponential
 factor function of only $k_{\perp}$ with Gaussian width $\langle k^2_{\perp} \rangle$
 and collinear pdf which is a function of both $x$ and 
 probing scale $Q$, seems to describe the experimental data (E288) of DY process \cite{E288}
 as shown in Ref. \cite{lowPT}. The evolution of $Q^2$ is taken only in  collinear pdf 
 which is known as Dokshitzer-Gribov-Lipatov-Altarelli-Parisi (DGLAP) or collinear evolution. 
However, Gaussian model was
proven to be unsuccessful to explain the high $P_T$ data of Z boson production in DY process
at CDF \cite{CDF} \cite{lowPT}. Moreover, Gaussian width depends on the energy of the experiment 
to describe $P_T$  distribution of high energy data \cite{lowPT}. In order to explain 
the high $P_T$ data one must look beyond DGLAP evolution. As mentioned
before, TMD factorization framework endows us evolution of TMDs \cite{tmde2}. By implementing the
evolution of TMDs, Ref. \cite{tmde3} has shown that  u quark TMD pdf has suppression at 
low $k_{\perp}$ 
and broad tail at high  $k_{\perp}$ values. The evolution of TMDs describes the high $P_{T}$ 
spectrum of Z and W boson productions perfectly \cite{tmde6}. It is very useful to study the 
evolution effect on $P_{T}$ distribution because the present as well as future experiments do 
operate at different energies.
In this work, the framework of Ref. \cite{tmde1} is used to implement TMD evolution.

In the recent past, a great attention has been paid to study charmonium and bottomonium 
productions as they provide QCD  of the formation of bound state. Here we introduce a clean 
promising process to estimate  the unpolarized gluon TMD $f^g_1(x, k_\perp)$ and
linearly polarized gluon distribution $h^{\perp g}_1(x, k_\perp)$ in $J/\psi$ and $\Upsilon$ 
production in unpolarized proton-proton collision i.e., $\text{pp}\rightarrow Q\overline{Q}+X$
 in the framework of transverse momentum dependent factorization. The partonic subprocesses
for charmonium and bottomonium production are the two gluon fusion process $gg\rightarrow Q\overline{Q}$ and 
$q\overline{q}\rightarrow Q\overline{Q}$, at leading order (LO).

Among existing three models for charmonium and bottomonium production, the first one is the color
singlet model (CSM) \cite{csm}. 
In CSM, the cross section for heavy quarkonium production is factorized  similar to QCD 
factorization theorem. The production
process can be decomposed into two steps. The first step is creation of on-shell heavy quark pair 
which is calculated perturbatively. The other step is the binding of the $Q\overline{Q}$ pair into physical 
color singlet state which is encoded in the long distance factor, the wave function. Generally 
the wave function
is obtained by fitting data or from potential models. CSM suggests that spin and color 
quantum numbers of $Q\overline{Q}$
pair do not alter during hadronization process. Therefore, to produce a physical color 
singlet state it requires that the   $Q\overline{Q}$ pair should be in color singlet state. 
Thus the name CSM for this model. Nevertheless, CSM is
unable to describe the large transverse momentum of $J/\psi$, $\Upsilon$ and $\psi(2s)$ 
\cite{csm1,csm2}.

The second one is NRQCD model \cite{com} in which  the cross section is  factorized  just like 
CSM. The quarkonium cross section is a product of short and long distance factors summed over 
all possible color, spin and angular momentum quantum numbers of the  $Q\overline{Q}$ pair. 
In this model the formation of heavy quark pair is either in color octet or color singlet state. 
The short distance factor can be calculated
with appropriate quantum numbers using perturbation theory. The long distance factor,
the  nonperturbative matrix element, describes the transition probability
of the $Q\overline{Q}$ pair from colored state to colorless physical state, which can be expanded 
in powers of $\upsilon$ where $\upsilon$ is the relative velocity of the heavy quark in
the quarkonium  rest frame. The values are $\upsilon^2=0.1$ and
$0.3$ for bottomonium and charmonium respectively.

The third model is the color evaporation model (CEM), which was first 
developed in 1977 by F.Halzen, Matsuda \cite{cem1} and Fritsch \cite{cem2}.
In CEM, it is assumed that  the heavy quark pair is produced perturbatively with definite spin 
and color quantum numbers which can be calculated upto  a desired order in $\alpha_{s}$. 
Thereafter, the heavy quark pair radiates soft gluons  to evolve into any physical color neutral 
quarkonium state with quantum numbers different than that of initial
heavy quark pair. The process of hadronization of the quarkonium from heavy quark pair is usually 
referred to as nonperturbative process. The CEM acquired the name 
\textquotedblleft color evaporation\textquotedblright, since the color 
of the initial $Q\bar{Q}$ pair does not affect the final
quarkonium state. According to CEM, the cross section of quarkonium state is some long distance
  factor times the cross section of the $Q\bar{Q}$ pair with invariant mass below the threshold mass.
 Typically, the long distance factors are considered to be  universal which are determined by
 fitting the heavy quark pair cross section with experimental data.
Schuler and R. Vogt \cite{cem5} determined the long distance factors for $J/\psi$ and 
$\Upsilon$ to be  0.055 and 0.087 respectively. Amundson $et~al$, \cite{cem6} introduced 
another constraint in CEM, namely the probability of producing
a color singlet quarkonium state is  only $1/9$ of the  heavy quark pair production. 
The long distance factors  deduced in this version of CEM for 
$J/\psi$ and $\Upsilon$ are 0.47 \cite{cem6} and 0.62 \cite{cem7} respectively. 
The general prediction of CEM is that the probability of forming any quarkonium state is 
independent of the color and spin quantum numbers. This model is found to be in close match with 
the experimental data with an inclusion of a phenomenological factor in the cross section 
that is dependent on the Gaussian distribution of the transverse momentum of the quarkonium 
\cite{cem3}. Given an importance to its simplicity,
the present work employs the CEM to advance the understanding of the TMDs
and their evolution. As a whole, the paper
contains five sections including introduction.
Section \ref{sec2} presents the formalism for $J/\psi$ and $\Upsilon$ production.
The  TMD evolution formalism is  presented in \ref{sec3}.
Numerical results are presented in the Section \ref{sec4} along with
the conclusion in Section \ref{sec5}.

\section{CHARMONIUM ($J/\psi$) AND BOTTOMONIUM ($\Upsilon$) PRODUCTION}\label{sec2}
The formalism for charmonium and bottomonium production in CEM is explained as follows. 
The cross section for this production is proportional to the rate of production of 
$ Q\overline{Q}$ pair which is integrated over the mass values ranging from $2m_Q$ to
$2m_{Q\overline{q}}$ \cite{cem4}, see also \cite{ours1,ours2} 
\be
\sigma=\frac{\rho}{9}\int_{2m_Q}^{2m_{Q\overline{q}}}dM\frac{d\sigma_{Q\overline{Q}}}{dM},
\ee
where $m_Q$ is the mass of charm or bottom quark and $m_{Q\overline{q}}$ is the mass of lightest
D meson for charmonium and B meson for bottomonium.  $\frac{d\sigma_{Q\overline{Q}}}{dM}$ can be
calculated perturbatively and $M$ is the invariant mass of the  $Q\overline{Q}$ pair.
 The probability of producing the quarkonium is zero if the invariant mass of heavy quark pair is
 more than $2m_{Q\overline{q}}$.$~$Here $\rho$ is long distance factor and we have taken 
 0.47 \cite{cem6} and 0.62 \cite{cem7} for production of $J/\psi$ and $\Upsilon$ respectively.
 \par
\begin{figure}[H]
\begin{center}
\includegraphics[width=8.5cm,height=4.0cm]{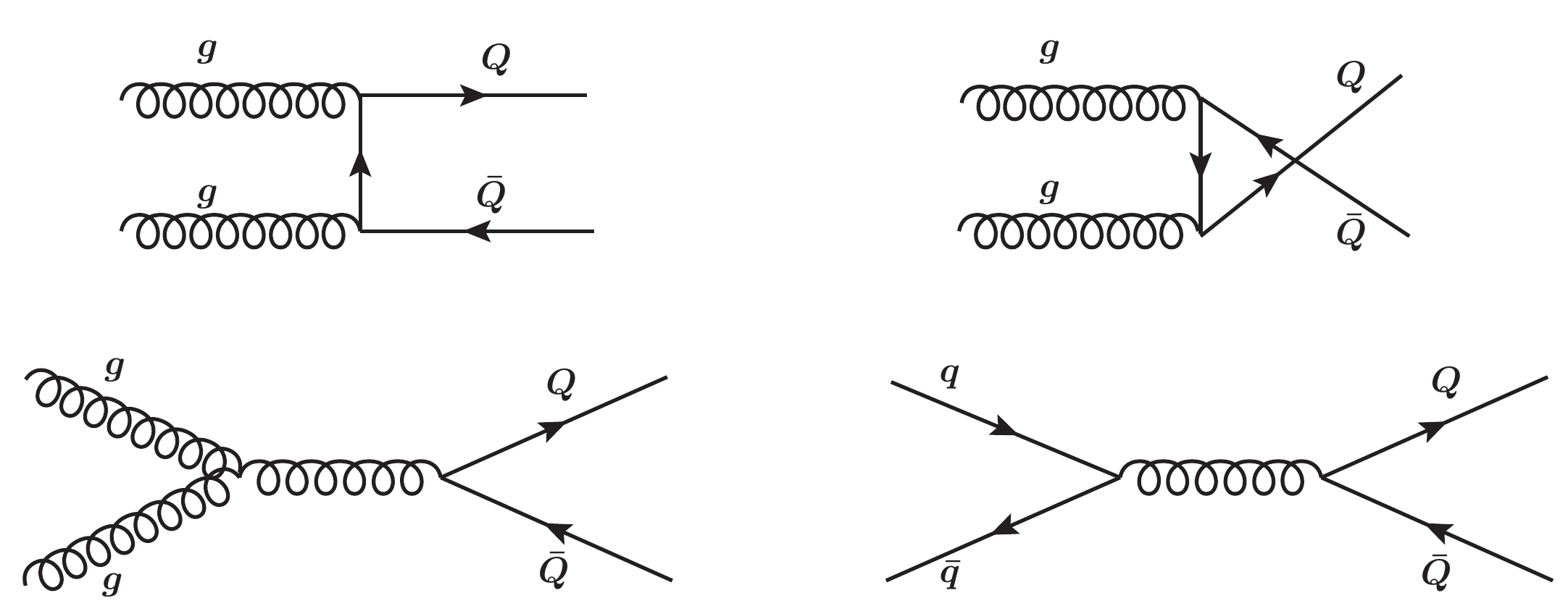}
\end{center}
\caption{\label{fig1} Feynman diagrams for gluon-gluon fusion and $q\bar{q}$ annihilation processes
at leading order. }
\end{figure}

 
 We consider the following process (unpolarized proton-proton collision) for charmonium and 
 bottomonium production
\be \label{eqp}
h(P_A)+h(P_B)\rightarrow Q\overline{Q}(q)+X,
\ee
where the four momenta are given within round brackets.
We choose the frame in which the proton $A$ is moving along $+\hat{z}$ axis and proton $B$ is moving along $-\hat{z}$ axis in
the center of mass (c.m) frame with four momenta $P^{\mu}_A=\frac{\sqrt{s}}{2}(1,0,0,1)$ and
$P_B^\mu=\frac{\sqrt{s}}{2}(1,0,0,-1)$ respectively. The $Q\overline{Q}$ pair is produced from 
two gluon fusion and $q\bar{q}$ annihilation partonic subprocesses, which are shown by Feynman 
diagrams in Fig.\ref{fig1}. The production of the quarkonium  is important only 
in the low transverse momentum region of the produced charmonium and bottomonium.
The assumption that the TMD factorization holds good at sufficiently high energies is adopted in
similar lines with Ref. \cite{prd57,prd660}. The cross section using generalized
CEM and TMD factorization formalism for Eq.\eqref{eqp} is
\begin{equation}\label{cross}
 \begin{aligned} 
 {d\sigma}={}&\frac{\rho}{9}\int dx_{a} dx_{b}  d^{2}{\bf k}_{\perp a} 
 d^2{\bf k}_{\perp b}\Bigg\{
 \Phi^{\mu\nu}_g(x_{a},{\bf k}_{\perp a})\Phi_{g\mu\nu}(x_{b},{\bf k}_{\perp b})
 {d\hat{\sigma}^{g g\rightarrow Q\overline{Q}}}+\\
&\Big[\Phi^q(x_{a},{\bf k}^2_{\perp a})
\Phi^{\bar{q}}(x_{b},{\bf k}^2_{\perp b})+
 \Phi^{\bar{q}}(x_{a},{\bf k}^2_{\perp a})\Phi^{q}(x_{b},{\bf k}^2_{\perp b})\Big]
 d\hat{\sigma}^{q\bar{q}\rightarrow Q\overline{Q}}\Bigg\},
\end{aligned}
\end{equation}
where $x_{a}$, $x_{b}$ are the longitudinal momentum fractions and $k_{\perp a}$, $k_{\perp b}$ 
are the transverse momentum of
the incoming gluons and quarks. Here, $\mathrm{q}=\mathrm{u,~\bar{u},~d,~\bar{d}},~s,~\bar{s}$ 
and $Q=\mathrm{c~or~b}$, depending on whether $J/\psi$ or $\Upsilon$ is
produced.
The gluon field strengths $F^{\mu\nu}(0)$ and $F^{\mu\nu}(\lambda)$ that are evaluated at fixed
light front time $\lambda^+=\lambda.n=0$ are used to define the gluon correlator 
(omitting gauge link) $\Phi^{\mu\nu}_g$. Here, $n$ is light-like vector and is conjugate to the
proton four momentum $\textit{P}$. At leading twist, the gluon correlator of unpolarized hadron contains two TMD gluon distribution
functions \cite{prd63}
\begin{equation}\label{gcorr}
 \begin{aligned} 
\Phi^{\mu\nu}_g(x,{\bf k}_{\perp})=&\frac{n_\rho n_\sigma}{({\textit {k}}.n)^2}\int 
\frac{d(\lambda .{\textit{ P}})d^2\lambda_T}{(2\pi)^3}
e^{ik.\lambda}\langle P|\mathrm{Tr}[F^{\mu\nu}(0)F^{\mu\nu}(\lambda)]|P\rangle|_{LF}\\
=&-\frac{1}{2x}\left\{g^{\mu\nu}_Tf^g_1(x,{\bf k}_{\perp}^2)-\left(\frac{k^{\mu}_{\perp}k^{\nu}_{\perp}}
{M^2_h}+g^{\mu\nu}_T\frac{{\bf k}^2_{\perp}}{2M_h^2}\right)
h^{\perp g}_1(x,{\bf k}_{\perp}^2)\right\}.
\end{aligned}
\end{equation}
The quark correlator with omitting the gauge link is defined  as \cite{jhep}
\begin{equation}\label{qcorr}
 \begin{aligned} 
\Phi^q(x,{\bf k}_{\perp})=&\int\frac{d(\lambda .{\textit{ P}})d^2\lambda_T}{(2\pi)^3}
e^{ik.\lambda}\langle P|\overline{\psi}(0)\psi(\lambda)]|P\rangle|_{LF}\\
=&\frac{1}{2}\left\{f^q_1(x,{\bf k}_{\perp}^2)\slashed{P}+ih^{\perp q}_1(x,{\bf k}_{\perp}^2)
\frac{[\slashed{k}_{\perp},\slashed{P}]}{2M_h}\right\},
\end{aligned}
\end{equation}
and the antiquark correlator is given by \cite{jhep}
\begin{equation}\label{qbarcorr}
 \begin{aligned} 
\Phi^{\bar{q}}(x,{\bf k}_{\perp})=&-\int\frac{d(\lambda .{\textit{ P}})d^2\lambda_T}{(2\pi)^3}
e^{-ik.\lambda}\langle P|\overline{\psi}(0)\psi(\lambda)]|P\rangle|_{LF}\\
=&\frac{1}{2}\left\{f^{\bar{q}}_1(x,{\bf k}_{\perp}^2)\slashed{P}+ih^{\perp {\bar{q}}}_1(x,{\bf k}_{\perp}^2)
\frac{[\slashed{k}_{\perp},\slashed{P}]}{2M_h}\right\}.
\end{aligned}
\end{equation}
Here $k^2_{\perp}=-{\bf k}^2_{\perp}$, $g^{\mu\nu}_T=g^{\mu\nu}-P^{\mu}n^{\nu}/P.n-n^{\mu}P^{\nu}/P.n$ and $M_h$ is the mass of 
hadron. The unpolarized and the linearly polarized gluon
distribution functions are respectively denoted by $f^g_1(x,{\bf k}_{\perp}^2)$ and
$h^{\perp g}_1(x,{\bf k}_{\perp}^2)$. Similarly,  $f^{q,\bar{q}}_1(x,{\bf k}_{\perp}^2)$ 
and  $h^{\perp q,\bar{q}}_1(x,{\bf k}_{\perp}^2)$ represent the distribution of 
unpolarized and transversely polarized  quark (antiquark) respectively.

The $d\hat{\sigma}$, the partonic cross section for $g g\rightarrow Q\overline{Q}$ and $q\bar{q}\rightarrow Q\overline{Q}$, is given by
\be
d\hat{\sigma}^{g g,~q\bar{q}\rightarrow Q\overline{Q}}=\frac{1}{2\hat{s}}\frac{d^3p_Q}{2E_Q}\frac{d^3p_{\bar{Q}}}{2E_{{\bar{Q}}}}
\frac{1}{(2\pi)^2}\delta^4(p_{a}+p_{b}-p_Q-p_{\bar{Q}})
\overline{|M_{gg,~q\bar{q}\rightarrow Q\bar{Q}}|^2},
\ee
where $p_{a}$ and $p_{b}$ are the four momentum vectors of incoming gluons and quarks and 
$p_{Q}$ $(p_{\bar{Q}})$ is the produced quark (antiquark) four momentum. Let us define a four
momentum vector $q=(q_0,{\bf q}_T,q_L)$ of the  $Q\bar{Q}$ pair, where $q_0$, $q_L$ and
${\bf q}_T$ are the energy, longitudinal and transverse components respectively.
The four momentum of quarkonium pair is $q=p_Q+p_{\bar{Q}}$ and using this relation one can
rewrite  $\frac{d^3p_{\bar{Q}}}{2E_{{\bar{Q}}}}=d^4p_{\bar Q}\delta(p^2_{\bar Q}-m^2_{\bar Q})$. By changing the variables $p_{\bar{Q}}$ and $q_0$ 
and $q_L$ to $q$ and $M^2$ and $y$ (rapidity) respectively \cite{prd67}, we get
\begin{equation} \label{cemm1}
 \begin{aligned}
 \frac{d^3p_{\bar{Q}}}{2E_{{\bar{Q}}}}&=d^4q\delta((q-p_Q)^2-m^2_Q),\\
dM^2dy&=2dq_0dq_L.
\end{aligned}
\end{equation}

The partonic cross section can be written as
\be\label{cemm2}
 \hat{\sigma}^{gg,~q\bar{q}\rightarrow Q\bar{Q}}=\frac{1}{2M^2}\int \frac{d^3p_Q}{2E_Q}
\delta((q-p_Q)^2-m^2_Q)
\overline{|M_{gg,~q\bar{q}\rightarrow Q\bar{Q}}|^2}.
\ee
The differential cross section Eq.\eqref{cross} can be written as functions of transverse momentum, rapidity and
squared mass of quarkonium using Eq.\eqref{cemm1} and \eqref{cemm2}
\begin{equation}\label{crosssection}
 \begin{aligned} 
 \frac{d^4\sigma}{dydM^2d^2{\bf q}_{T}}={}&\frac{\rho}{18}\int dx_{a} dx_{b}  
 d^{2}{\bf k}_{\perp a} d^2{\bf k}_{\perp b}\delta^4(p_{a}+p_{b}-q)\Bigg\{
 \Phi^{\mu\nu}_g(x_{a},{\bf k}_{\perp a})\Phi_{g\mu\nu}(x_{b},{\bf k}_{\perp b})
 {\hat{\sigma}^{g g\rightarrow Q\overline{Q}}}+\\
&\Big[\Phi^q(x_{a},{\bf k}^2_{\perp a})
\Phi^{\bar{q}}(x_{b},{\bf k}^2_{\perp b})+
 \Phi^{\bar{q}}(x_{a},{\bf k}^2_{\perp a})\Phi^{q}(x_{b},{\bf k}^2_{\perp b})\Big]
 \hat{\sigma}^{q\bar{q}\rightarrow Q\overline{Q}}\Bigg\}.
\end{aligned}
\end{equation}
To obtain the differential cross section in terms of TMD  distribution functions, we substitute 
Eq.\eqref{gcorr}, \eqref{qcorr} and \eqref{qbarcorr} in Eq.\eqref{crosssection} 
\begin{equation}
 \begin{aligned}
\frac{d^4\sigma}{dydM^2d^2{\bf q}_{T}}={} &\frac{\rho}{18}\int dx_{a}dx_{b}d^{2}{\bf k}_{\perp a} d^2{\bf k}_{\perp b}
\delta^4(p_{a}+p_{b}-q)\Bigg\{\frac{1}{2x_ax_b}    \\
&\times \left[f_1^g(x_{a},{\bf k}_{\perp a}^2)f_1^g(x_{b},{\bf k}_{\perp b}^2) 
 +wh_1^{\perp g}(x_{a},{\bf k}_{\perp a}^2)h_1^{\perp g}(x_{b},{\bf k}_{\perp b}^2)\right]
\hat{\sigma}^{g g\rightarrow Q\overline{Q}}(M^2)\\
& +\frac14\sum_\mathrm{q}\Big[f_1^q(x_{a},{\bf k}^2_{\perp a})
f_1^{\bar{q}}(x_{b},{\bf k}^2_{\perp b})+
 f_1^{\bar{q}}(x_{a},{\bf k}^2_{\perp a})f_1^{q}(x_{b},{\bf k}^2_{\perp b})\Big]
 \hat{\sigma}^{q\bar{q}\rightarrow Q\overline{Q}}(M^2)\Bigg\},
 \end{aligned}
\end{equation}
where $w$ is weight factor of the transverse momentum 
\be
w=\frac{1}{2M_h^4}\left[({\bf k}_{\perp a}.{\bf k}_{\perp b})^2-\frac{1}{2}{\bf k}_{\perp a}^2{\bf k}_{\perp b}^2\right].
\ee
In most parts of this study, the transversely polarized quark and antiquark contributions have been neglected since 
 the gluon channel is dominant, as shown later.
The total partonic cross sections $\hat{\sigma}$ are calculated perturbatively \cite{pl79}
\be
 \hat{\sigma}^{g g\rightarrow Q\bar{Q}}&=&\frac{\pi\alpha^2_s}{3M^2}
 \left[ \left(1+\gamma+\frac{1}{16}\gamma^2\right)\ln\frac{1+\sqrt{1-\gamma}}{1-\sqrt{1-\gamma}}-
 \left(\frac{7}{4}+\frac{31}{16}\gamma\right)\sqrt{1-\gamma}\right],\\
 \hat{\sigma}^{q\bar{q}\rightarrow Q\bar{Q}}&=&\frac{2}{9}\left(\frac{4\pi\alpha^2_s}{3M^2}\right)
 \left(1+\frac{1}{2}\gamma\right)\sqrt{1-\gamma},
\ee
 where $\gamma=\frac{4m^2_Q}{M^2}$ and $M^2=\hat{s}$, $\sqrt{\hat{s}}$ is
the center-of-mass energy of the partonic subprocess.  
In line with Ref. \cite{prd67}, the four momentum conservation $\delta$ function can also be written as
 \be
 \delta^4(p_{a}+p_{b}-q)&=&\delta(E_{a}+E_{b}-q_0)\delta(p_{za}+p_{zb}-q_L)
  \delta^2({\bf k}_{\perp a}+{\bf k}_{\perp b}-{\bf q}_T)\\
  &=&\frac{2}{s}\delta\left(x_{a}-\frac{Me^{y}}{\sqrt{s}} \right) \label{delta}
  \delta\left(x_{b}-\frac{Me^{-y}}{\sqrt{s}} \right)
  \delta^2({\bf k}_{\perp a}+{\bf k}_{\perp b}-{\bf q}_T).
\ee
After performing  integrations over $x_{a}$ and $x_{b}$, the two $\delta$ functions in Eq.\eqref{delta} gives
\be
x_{a,b}=\frac{M}{\sqrt{s}}e^{\pm y}.
\ee
$\sqrt{s}$ is the center-of-mass energy. The expression for the cross section obtained is as follows
\begin{equation}
\begin{aligned}
\frac{d^4\sigma}{dydM^2d^2{\bf q}_{T}}={} &\frac{\rho}{9s}\int d^2{\bf k}_{\perp a} d^2{\bf k}_{\perp b}
\delta^2({\bf k}_{\perp a}+{\bf k}_{\perp b}-{\bf q}_T)\Bigg\{\frac{1}{2}
\Big[f_1^g(x_{a},{\bf k}_{\perp a}^2)f_1^g(x_{b},{\bf k}_{\perp b}^2) \\
& +wh_1^{\perp g}(x_{a},{\bf k}_{\perp a}^2)h_1^{\perp g}(x_{b},{\bf k}_{\perp b}^2)\Big]
\hat{\sigma}^{g g\rightarrow Q\overline{Q}}(M^2)\\
&+\frac14\sum_\mathrm{q}\Big[f_1^q(x_{a},{\bf k}^2_{\perp a})
f_1^{\bar{q}}(x_{b},{\bf k}^2_{\perp b})+
 f_1^{\bar{q}}(x_{a},{\bf k}^2_{\perp a})f_1^{q}(x_{b},{\bf k}^2_{\perp b})\Big]
 \hat{\sigma}^{q\bar{q}\rightarrow Q\overline{Q}}(M^2)\Bigg\}.
\end{aligned}
\end{equation}
We can eliminate ${\bf k}_{\perp b}$ by integrating and finally we reach
\be
\frac{d^2\sigma^{ff+hh}}{dydq^2_T}=\frac{d^2\sigma^{ff}}{dydq^2_T}+\frac{d^2\sigma^{hh}}{dydq^2_T},
\ee
where
\begin{equation}
\begin{aligned}
\frac{d^2\sigma^{ff}}{dydq^2_T}={} &\frac{\rho}{36s} \int dM^2\int d\phi_{q_T}\int d{ k}_{\perp a} k_{\perp a}
\int d\phi_{k_{\perp a}}\Bigg\{
f_1^g(x_{a},{\bf k}_{\perp a}^2)\\
&\times f_1^g(x_{b},({\bf q}_{T}-{\bf k}_{\perp a})^2)
\hat{\sigma}^{g g\rightarrow Q\overline{Q}}(M^2)+
\frac12\sum_q\Big[f_1^q(x_{a},{\bf k}^2_{\perp a})
f_1^{\bar{q}}(x_{b},({\bf q}_{T}-{\bf k}_{\perp a})^2)\\
&+ f_1^{\bar{q}}(x_{a},{\bf k}^2_{\perp a})f_1^{q}(x_{b},({\bf q}_{T}-{\bf k}_{\perp a})^2)\Big]
 \hat{\sigma}^{q\bar{q}\rightarrow Q\overline{Q}}(M^2)\Bigg\},
\end{aligned}
\end{equation}
and
\begin{equation}
\begin{aligned}
\frac{d^2\sigma^{hh}}{dydq^2_T}={} &\frac{\rho}{36s}\frac{1}{2M_h^4} \int dM^2\int d\phi_{q_T}\int d{ k}_{\perp a} k_{\perp a}
\int d\phi_{k_{\perp a}}\\
& \times\left[ \frac{1}{2}k^4_{\perp a}-\frac{1}{2}k^2_{\perp a}q^2_T 
 - q_T k^3_{\perp a}\cos(\phi_{k_{\perp a}}-\phi_{q_T})+q^2_T
 k^2_{\perp a}\cos^2(\phi_{k_{\perp a}}-\phi_{q_T}) \right]\\
&\times h_1^{\perp g}(x_{a},{\bf k}_{\perp a}^2)h_1^{\perp g}(x_{b},({\bf q}_{T}-{\bf k}_{\perp a})^2)
\hat{\sigma}^{g g\rightarrow Q\overline{Q}}(M^2).
\end{aligned}
\end{equation}

Here, $\phi_{k_{\perp a}}$ is the azimuthal angle of gluon and quark and the azimuthal 
angle of the quarkonium is $\phi_{q_T}$.

\section{TMD Evolution}\label{sec3}
In the present section, we discuss the model used for the TMDs as well as
the TMD evolution. As per the general conception, we assume that unpolarized distribution 
functions of gluons, quarks and
antiquarks TMDs do simply depend on Gaussian form of gluon's and quark's 
transverse momentum \cite{prd72}
\be
 f_1^{g,~q}(x,{\bf k}^2_{\perp })=f_1^{g,~q}(x,Q^2)\frac{1}{\pi \langle k^2_{\perp }\rangle}
 e^{-{\bf k}^2_{\perp }/\langle k^2_{\perp }\rangle}.
\ee
Here, TMD pdf is factorized into  $k_{\perp }$ and $x$ dependencies. $Q^2$ dependence is only  
in $f_1^{g, ~q}(x,Q^2)$ which is the usual collinear pdf evaluated at scale $Q^2$.
We have chosen $Q^2=M^2$ that is known as collinear or DGLAP evolution. The
factorized form of $h_1^{\perp g}$ \cite{prd86} is given by
\be\label{hg}
h_1^{\perp g}(x,{\bf k}^2_{\perp })=\frac{M^2_hf_1^g(x,Q^2)}{\pi\langle k^2_{\perp }\rangle^2}\frac{2(1-r)}{r}e^{1-
 {\bf k}^2_{\perp }\frac{1}{r\langle k^2_{\perp }\rangle}},
\ee
where $r$ is the parameter which has the range $0<r<1$. The Eq.\eqref{hg} obeys the model 
independent positive bound \cite{prl106} for all values of $x$ and $k_{\perp }$
\be
\frac{{\bf k}_{\perp}^2}{2M_h^2}|h^{\perp g }_1(x,{\bf k}_{\perp }^2)|\leq 
f_1^g(x,{\bf k}^2_{\perp }).
\ee
In this work, we use two values for squared intrinsic average transverse momentum of gluons and quarks i.e., 
$\langle k^2_{\perp }\rangle=0.25$ GeV$^2$ and 1 GeV$^2$ \cite{prd86}. 
The parameter values chosen are $r=\frac13$ and $\frac23$  \cite{prd86}.

\subsection{Model-I}\label{m1}
In model-I, we do not use an upper limit of the transverse momentum
integration. The exponential behavior of unpolarized and linearly polarized TMDs allows 
us to integrate 
analytically with respect to $k_{\perp a}$ and we 
obtain 
\be\label{m1eq1}
\frac{d^2\sigma^{ff}}{dydq^2_{T}} =\frac{\beta\rho}{36s}\int^{{4m^2_{Q\bar{q}}}}_{4m^2_Q}dM^2
e^{-q^2_T\frac{\beta}{2}}\Bigg\{  f_1^g(x_{a}) f_1^g(x_{b}) 
 \hat{\sigma}^{g g\rightarrow Q\overline{Q}}(M^2)\nonumber\\
 +\frac12\sum_q \Big[f_1^q(x_{a})f_1^{\bar{q}}(x_{b})+
 f_1^{\bar{q}}(x_{a})f_1^{q}(x_{b})\Big]
 \hat{\sigma}^{q\bar{q}\rightarrow Q\overline{Q}}(M^2)\Bigg\},
\ee
and
\be\label{m1eq2}
 \frac{d^2\sigma^{hh}}{dydq^2_T}=  \frac{\beta\rho r(1-r)^2}{72s}\int^{{4m^2_{Q\bar{q}}}}_{4m^2_Q} dM^2 
\left[1-\frac{\beta q^2_T}{r}+\frac{\beta^2q^4_T}{8r^2}\right]
e^{[2-\frac{\beta}{2r}q^2_T]}  f_1^g(x_{a}) f_1^g(x_{b})\nonumber\\
\times \hat{\sigma}^{g g\rightarrow Q\overline{Q}}(M^2),
\ee

where $\beta=\frac{1}{\langle k^2_{\perp a}\rangle}=\frac{1}{\langle(q_T- k_{\perp a})^2\rangle}$.
\subsection{Model-II}\label{m2}
In this model, we consider the effective intrinsic motion of the gluons for 
Gaussian distribution to be restricted to $k_{\mathrm{max}}=\sqrt{\langle {k}^2_{\perp a}\rangle}$ 
\cite{epj39}. Hence the expressions for quarkonium production are given by
\begin{equation}\label{m2eq1}
\begin{aligned}
\frac{d^2\sigma^{ff}}{dydq^2_T} =& \frac{\beta^2\rho}{36s\pi^2}\int^{{4m^2_{Q\bar{q}}}}_{4m^2_Q}
dM^2\int d\phi_{q_T}
 \int_0^{k_{\mathrm{max}}} dk_{\perp a}k_{\perp a}\int d\phi_{k_{\perp a}} e^{-\Delta\beta}\\
 \times&\Bigg\{ f_1^g(x_{a}) f_1^g(x_{b}) \hat{\sigma}^{g g\rightarrow Q\overline{Q}}(M^2)+
 \frac12\sum_q\Big[f_1^q(x_{a})f_1^{\bar{q}}(x_{b})+ f_1^{\bar{q}}(x_{a})f_1^{q}(x_{b})\Big]\\
&\times\hat{\sigma}^{q\bar{q}\rightarrow Q\overline{Q}}(M^2)\Bigg\},
 \end{aligned}
\end{equation}
and 
\begin{equation}\label{m2eq2}
\begin{aligned}
 \frac{d^2\sigma^{hh}}{dydq^2_T}={} &\frac{\beta^4\rho(1-r)^2}{18sr^2\pi^2}\int^{{4m^2_{Q\bar{q}}}}_{4m^2_Q} 
 dM^2\int d\phi_{q_T} \int_0^{k_{\mathrm{max}}} d{ k}_{\perp a} k_{\perp a} \int d\phi_{k_{\perp a}}   \\
\times &\left[ \frac{1}{2}k^4_{\perp a}-\frac{1}{2}k^2_{\perp a}q^2_T 
 - q_T k^3_{\perp a}\cos(\phi_{k_{\perp a}}-\phi_{q_T})+q^2_T
 k^2_{\perp a}\cos^2(\phi_{k_{\perp a}}-\phi_{q_T}) \right]  \\ 
&\times e^{[2-\frac{\beta}{r}\Delta]}
  f_1^g(x_{a}) f_1^g(x_{b})
 \hat{\sigma}^{g g\rightarrow Q\overline{Q}}(M^2),
 \end{aligned}
\end{equation}

where $\Delta=2k^2_{\perp a}+q^2_T-2q_Tk_{\perp a}\cos(\phi_{k_{\perp a}}-\phi_{q_T})$.

For the evolution of  TMDs we adopted the formalism Ref. \cite{tmde1} as mentioned in the 
introduction. The TMD evolution formalism has been formulated in two dimensional coordinate 
space ($b_{\perp}$-space). Therefore,
transverse momentum dependent gluon-gluon correlator function is Fourier transformed 
into $b_\perp$-space which is defined as
\be\label{et1}
\Phi(x,{\bf b}_\perp)=\int d^2{\bf k}_\perp e^{-i{\bf k}_{\perp}.{\bf b}_{\perp}}\Phi(x,{\bf k}_\perp),
\ee
and the inverse Fourier transformation is 
\be\label{et2}
\Phi(x,{\bf k}_\perp)=\frac{1}{(2\pi)^2}\int d^2{\bf b}_\perp 
e^{i{\bf k}_{\perp}.{\bf b}_{\perp}}\Phi(x,{\bf b}_\perp).
\ee
After performing delta function integrations in  Eq.\eqref{crosssection}, the differential cross 
section of quarkonium  can be written as following
\begin{equation}\label{kcrosssection}
 \frac{d^4\sigma}{dydM^2d^2{\bf q}_{T}}=\frac{\rho}{9s}\int  d^{2}{\bf k}_{\perp a}
 \Phi^{\mu\nu}_g(x_{a},{\bf k}_{\perp a})\Phi_{g\mu\nu}(x_{b},{\bf q}_T-{\bf k}_{\perp a})
 \hat{\sigma}^{g g\rightarrow Q\overline{Q}}.
\end{equation}
In the evolution of TMDs, only TMD pdfs of gluons are considered. Quark 
contribution is neglected because of its insignificance in the quarkonium production as depicted
in Fig. \ref{fig3}.
Substituting Eq.\eqref{et1} and \eqref{et2} in \eqref{kcrosssection}, one can  obtain the following
differential cross section in $b_\perp$-space as,
\begin{equation}\label{bcrosssection}
 \frac{d^4\sigma}{dydM^2d^2{\bf q}_{T}}=\frac{\rho}{9s}
 \frac{1}{(2\pi)^2}\int d^2{\bf b}_\perp e^{i{\bf q}_{T}.{\bf b}_{\perp}}
 \Phi^{\mu\nu}_g(x_{a},{\bf b}_{\perp})\Phi_{g\mu\nu}(x_{b},{\bf b}_{\perp })
 \hat{\sigma}^{g g\rightarrow Q\overline{Q}}.
\end{equation}
In lines of Ref. \cite{tmde1}, the gluon correlator function in $b_\perp$-space is given by
\be\label{et3}
\Phi^g(x,{\bf b}_\perp)=\frac{1}{2x}\left\{g^{\mu\nu}_Tf^g_1(x,{\bf b}_{\perp}^2)-
\left(\frac{2b^{\mu}_{\perp}b^{\nu}_{\perp}}
{b^2_{\perp}}-g^{\mu\nu}_T\right)h^{\perp g}_1(x,{\bf b}_{\perp}^2)\right\}.
\ee
The differential cross section of quarkonium in terms of  $b_\perp$-space pdfs is obtained by 
inserting Eq.\eqref{et3} in Eq.\eqref{bcrosssection}
\begin{equation}\label{et4}
 \begin{aligned} 
 \frac{d^4\sigma}{dydM^2d^2{\bf q}_T}={}&\frac{\rho}{18s}\frac{1}{2\pi}
 \int_0^{\infty}b_{\perp} db_{\perp}J_0(q_Tb_{\perp})
 \Big\{ f_1^g(x_{a}, b_{\perp}^2)f_1^g(x_{b}, b_{\perp}^2)\\ 
& +h_1^{\perp g}(x_{a}, b_{\perp}^2)h_1^{\perp g}(x_{b},b_{\perp}^2)\Big\}
  {\hat{\sigma}^{g g\rightarrow Q\overline{Q}}(M^2)},
\end{aligned}
\end{equation}
where $J_0$ is the Bessel function of zeroth order.
TMD pdfs depend not only on the renormalization scale $\mu$ but also on $\zeta$. 
Here, $\zeta$  is an auxiliary parameter which is introduced to regularize the light 
cone divergence in TMD factorization formalism \cite{tmde2}.
 Collins-Soper (CS) and Renormalization Group (RG) equations are obtained by taking 
evolution in $\zeta$ and $\mu$  
 respectively \cite{tmde2,tmde3}. Using CS and RG equations one obtains the
 evolution of TMDs from initial
 scale $Q_i=c/b_{\ast}(b_{\perp})$ to  final scale $Q_f=Q$
 \cite{tmde3,tmde4}:
  \begin{eqnarray}{\label{pert}}
  f(x,b_{\perp},Q_f,\zeta)=f(x,b_\perp,Q_i,\zeta)R_{pert}\left(Q_f,Q_i,b_{\ast}\right)
  R_{NP}\left(Q_f,Q_i,b_{\perp}\right),
 \end{eqnarray}
 where $R_{pert}$ and $R_{NP}$ are perturbative and non-perturbative parts
 respectively. $c/b_{\ast}$
 is the initial scale where $c=2e^{-\gamma_\epsilon}$ bearing Euler's 
 constant $\gamma_\epsilon\approx0.577$. Here,
 $b_{\ast}(b_{\perp})=\frac{b_{\perp}}{\sqrt{1+\left(\frac{b_{\perp}}{b_{\mathrm{max}}}\right)^2}}\approx b_{\mathrm{max}}$
 when $b_{\perp}\rightarrow \infty$ and $b_{\ast}(b_{\perp})\approx b_{\perp}$ when $b_{\perp}\rightarrow 0$ is usually 
 known as $b_{\ast}$ prescription. This prescription is used to separate out nonperturbative 
part from the evolution kernel 
 since the evolution kernel is not valid at larger values of $b_{\perp}$ \cite{tmde2}. 
The separated nonperturbative part is 
 embodied in the exponential containing the nonperturbative Sudakov factor, $R_{NP}$.
 The  evolution kernel  is  given by \cite{tmde1}
 \be\label{sudakov}
 R_{pert}\left(Q_f,Q_i,b_{\ast}\right)=\mathrm{exp}\Big\{{-\int_{c/b_{\ast}}^{Q}
\frac{d\mu}{\mu}\left(A\log\left(\frac{Q^2}
 {\mu^2}\right)+B\right)}\Big\},
 \ee
 where $A$ and $B$ are anomalous dimensions of evolution kernel and TMD pdf respectively which have perturbative expansion
  like $$A=\sum_{n=1}^{\infty}\left(\frac{\alpha_s(\mu)}{\pi}\right)^nA_n$$
 and $$B=\sum_{n=1}^{\infty}\left(\frac{\alpha_s(\mu)}{\pi}\right)^nB_n.$$ 
 The first order expansion  coefficients in $\alpha_s$ are $A_1=C_A$ and $B_1=-\frac{1}{2}
 (\frac{11}{3}C_A-\frac{2}{3}N_f)$. The anomalous dimensions are derived up to 
3-loop level \cite{tmde5}. The perturbative Sudakov factor, in our case, is
the same for unpolarized and linearly polarized gluon TMDs
\cite{tmde7,tmde6}.
 The evolution kernel resummed up to NLL accuracy in exploration of Sivers asymmetry in SIDIS and  DY processes \cite{tmde6}. Generally, 
 the nonperturbative
 factor, $R_{NP}$, is extracted by fitting with experimental data. We choose the nonperturbative 
 Sudakov factor as given in Aybat et al. \cite{tmde3} which describes the SIDIS and Z boson 
data with good accuracy.
 \be
R_{NP}=\mathrm{exp}\left\{-\left[\frac{g_2}{2}\log\frac{Q}{2Q_0}+\frac{g_1}{2}\left(1+2g_3\log\frac{10xx_0}{x_0+x}\right)
\right]b_{\perp}^2\right\},
 \ee
 where the best fit parameters are \cite{tmde1} 
 \be
 g_1=0.201\mathrm{~GeV^2},~~~~~~~~g_2=0.184\mathrm{~GeV^2},~~~~~~~~~~~g_3=-0.129,\nonumber\\
 Q_0=1.6\mathrm{~GeV},~~~~b_{\mathrm{max}}=1.5\mathrm{~GeV^{-1}},~~~~x_0=0.009,~~~~x=0.09.
 \ee
 So far no experimental data is available to extract the nonperturbative fitting parameters of 
 linearly polarized gluon TMD. Hence the same nonperturbative Sudakov factor is chosen for
 linearly polarized gluon TMD pdf which is considered for unpolarized distribution function.
 However, as discussed in \cite{tmde1}, the $Q$ independent part of the
 nonperturbative Sudakov factor is expected to depend on spin, so one should
 in principle use a different $Q$-independent part for the linearly
 polarized gluon distribution; but the difference does not affect the result
 at large $Q$. The TMD  distribution function  $f(x,b_\perp,Q_i,\zeta)$ 
 is formally written in terms of a product of convolution 
 of coefficient function and standard   collinear pdf \cite{tmde3} 
  \be
  f(x,b_\perp,Q_i,\zeta)=\sum_{i=g,q}\int_x^1\frac{d\hat{x}}{\hat{x}}C_{i/g}(x/\hat{x},b_{\perp},\alpha_s,\mu,\zeta)
  f_{i/p}(\hat{x},c/b_{\ast})+\mathcal{O}(b_{\perp}\varLambda_{QCD}).
  \ee
  The coefficient function is calculated perturbatively which is different for each 
TMD pdf and independent of process. The collinear pdf produces the perturbative tail at 
small $b_{\perp}$ values which is evaluated at scale
$c/b_{\ast}$ rather than $Q$ in contrast to the  DGLAP evolution. We neglect quark contribution
 since it's effect is small compared to gluon. The unpolarized and linearly polarized TMD pdf at leading and first order 
 in $\alpha_s$ are given by \cite{tmde1}
\be\label{et5}
  f_1^g(x,b_\perp,Q_i,\zeta)=f_{g/p}(x,c/b_{\ast})+\mathcal{O}(\alpha_s),
 \ee
 \be\label{et6}
 h_1^{\perp g}(x,b_\perp,Q_i,\zeta)=\frac{\alpha_s(c/b_{\ast})C_A}{\pi}\int_x^1
 {d \hat{x}\over \hat{x}}\left(\frac{\hat{x}}{x}-1\right)
 f_{g/p}(\hat{x},c/b_{\ast})+\mathcal{O}(\alpha_s^2).
 \ee
 Now we can write the Eq.\eqref{et4} as the following by using Sudakov factors and TMD pdfs
 \be \label{tmdevo}
\frac{d^2\sigma^{ff+hh}}{dydq^2_T}=\frac{d^2\sigma^{ff}}{dydq^2_T}+\frac{d^2\sigma^{hh}}{dydq^2_T},
\ee
where
 \begin{equation}\label{evoleq1}
\begin{aligned}
 \frac{d^2\sigma^{ff}}{dydq^2_T}={} &\frac{\rho}{36s}\int^{{4m^2_{Q\bar{q}}}}_{4m^2_Q}dM^2
 \int_0^{\infty}b_{\perp} db_{\perp}J_0(q_Tb_{\perp})
 f^g_1(x_a,c/b_{\ast})f_1^g(x_b,c/b_{\ast})
 {\hat{\sigma}^{g g\rightarrow Q\overline{Q}}(M^2)}\\
&\mathrm{exp}\Bigg\{{-2\int_{c/b_{\ast}}^{Q}\frac{d\mu}{\mu}\left(A\log\left(\frac{Q^2}
 {\mu^2}\right)+B\right)}\Bigg\}\mathrm{exp}\Bigg\{-\Big[0.184\log\frac{Q}{2Q_0}
 +0.332\Big]b_{\perp}^2\Bigg\},
 \end{aligned}
\end{equation}
and 
\begin{equation}\label{evoleq2}
\begin{aligned}
 \frac{d^2\sigma^{hh}}{dydq^2_T}={} &\frac{\rho C_A^2}{36s\pi^2}
 \int^{{4m^2_{Q\bar{q}}}}_{4m^2_Q}dM^2
 \int_0^{\infty}b_{\perp} db_{\perp}J_0(q_Tb_{\perp})\alpha_s^2(c/b_{\ast})
 {\hat{\sigma}^{g g\rightarrow Q\overline{Q}}(M^2)}\\
& \int_{x_a}^1\frac{dx_1}{x_1}\left(\frac{x_1}{x_a}-1\right)f^g_1(x_1,c/b_{\ast})
\int_{x_b}^1\frac{dx_2}{x_2}\left(\frac{x_2}{x_b}-1\right)f_1^g(x_2,c/b_{\ast})\\
&\mathrm{exp}\Bigg\{{-2\int_{c/b_{\ast}}^{Q}
\frac{d\mu}{\mu}\left(A\log\left(\frac{Q^2}
 {\mu^2}\right)+B\right)}\Bigg\}\mathrm{exp}\Bigg\{-\Big[0.184\log\frac{Q}{2Q_0}
 +0.332\Big]b_{\perp}^2\Bigg\}.
 \end{aligned}
\end{equation}
 \section{Numerical Results} \label{sec4}
In this section we present our numerical results.
As for CEM, the squared invariant mass 
of the quark pair is integrated from $4m^2_Q$ to $4m^2_{Q\bar{q}}$. For $J/\psi$ production, 
we have taken charm quark mass ($m_c=1.275$ GeV) for $m_Q$ and lightest D meson mass ($m_D=1.863$ GeV) for $m_{Q\bar{q}}$. 
Moreover, bottom quark mass ($m_b=4.18$ GeV) for  $m_Q$ and lightest B meson mass ($m_B=5.279$ GeV) for $m_{Q\bar{q}}$
have been considered for  $\Upsilon$ production. MSTW2008 \cite{mstw} has been used for pdfs to obtain
the differential cross section.\par
We have calculated the LO transverse momentum ($q_T$) and rapidity ($y$) distributions of $J/\psi$ and  
$\Upsilon$, we also present the cross section differential in $q_T$ using TMD evolution 
formalisms at  center-of mass energies of LHCb ($\sqrt{s}=7$ TeV), RHIC ($\sqrt{s}=500$ GeV) and AFTER 
($\sqrt{s}=115$ GeV) experiments.
The rapidity of quarkonium is integrated in the range of $y\in[2.0,4.5]$, $y\in[-3.0,3.0]$ and 
 $y\in[-0.5,0.5]$ for LHCb, RHIC and AFTER respectively, to obtain the differential 
cross section as a function of $q_T$. The rapidity distribution of quarkonium has been 
calculated by integrating $q_T$ from 0 to 0.5 GeV  for all energies. 
 The conventions in the figures are the following. In all the figures, 
\textquotedblleft ff\textquotedblright~ denotes contributions of unpolarized
TMDs only in the cross section and \textquotedblleft ff+hh\textquotedblright~
means both unpolarized pdfs and linearly polarized gluon distributions are
taken into account.

\begin{figure}[H]
\begin{minipage}{0.99\textwidth}
\includegraphics[width=7.5cm,height=6cm]{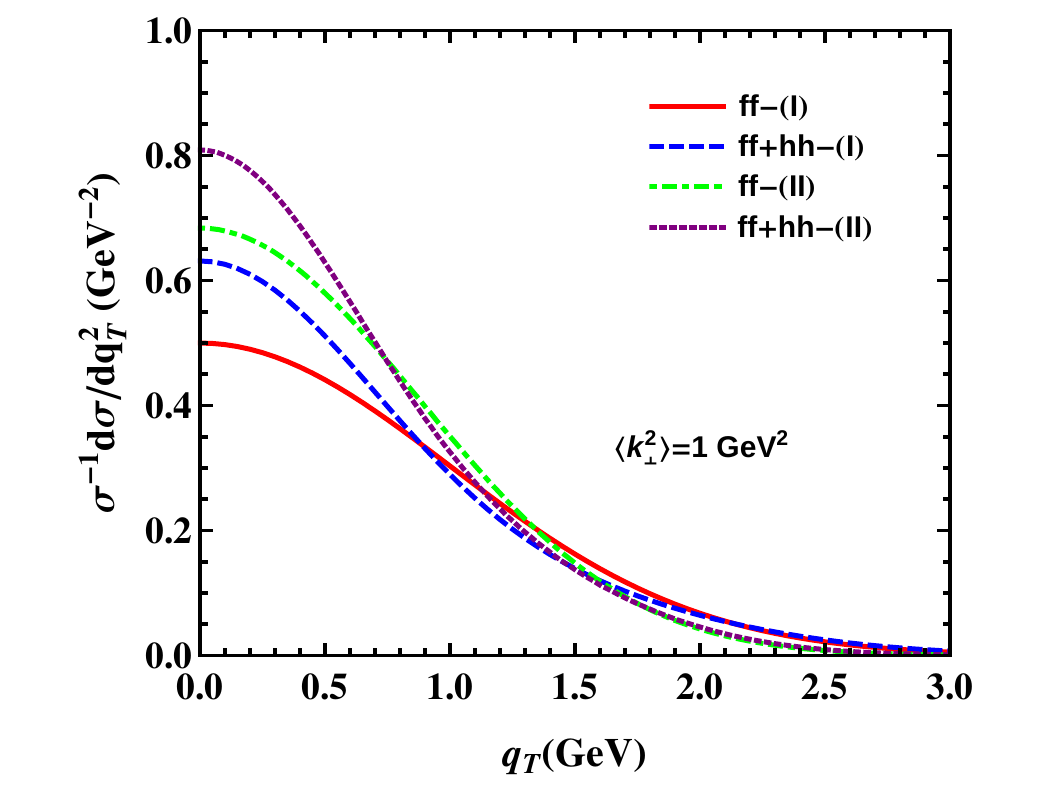}
\hspace{0.1cm}
\includegraphics[width=7.5cm,height=6cm]{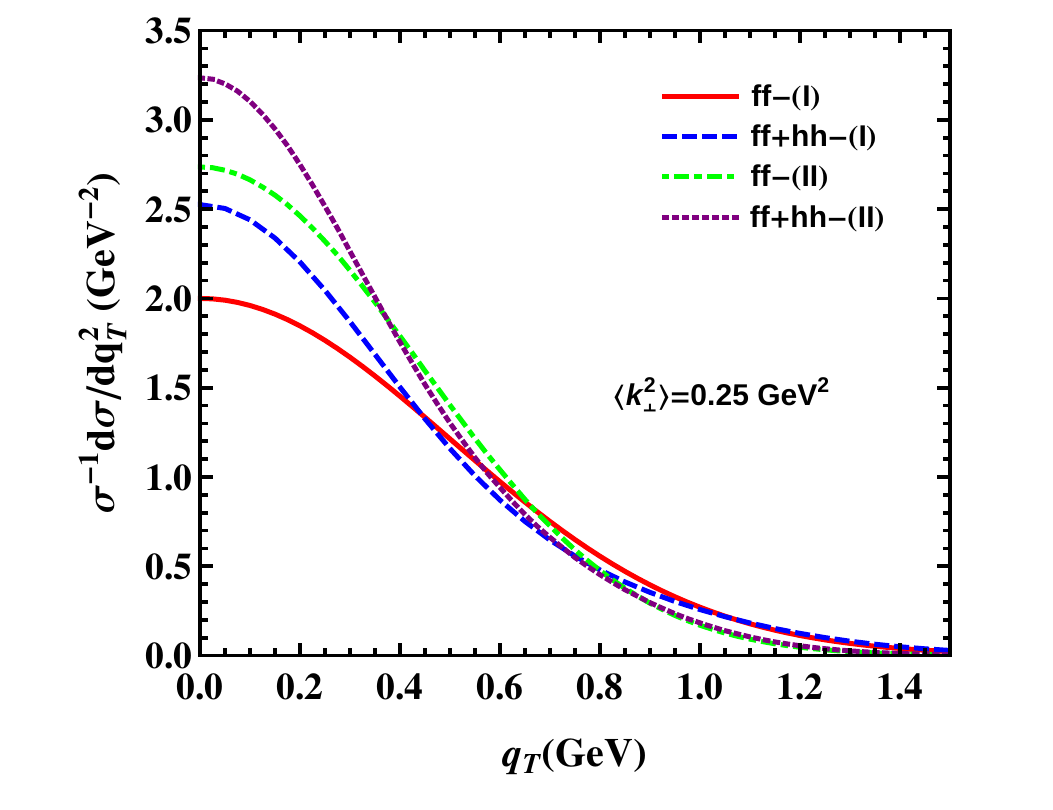}
\end{minipage}
\caption{\label{fig2}(color online) Differential cross section (normalized) of 
 $J/\psi$  and $\Upsilon$ production in $\text{pp}\rightarrow Q\overline{Q}+X$  
at LHCb ($\sqrt{s}=7$ TeV),  RHIC ($\sqrt{s}=500$ GeV) and AFTER ($\sqrt{s}=115$ GeV) energies
using DGLAP evolution approach  For  $r=\frac{2}{3}$  .
The solid (ff-(I)) and dot dashed (ff-(II)) lines are obtained by considering
unpolarized gluons and quarks in Model-I and Model-II respectively.
The dashed (ff+hh-(I)) and tiny dashed (ff+hh-(II)) lines are obtained by  taking 
into account unpolarized gluons and quarks plus linearly polarized gluons in Model-I and Model-II
respectively. See the text for ranges of rapidity integration.}
\end{figure}

\begin{figure}[H]
\begin{minipage}{0.99\textwidth}
\includegraphics[width=7.5cm,height=6cm,clip]{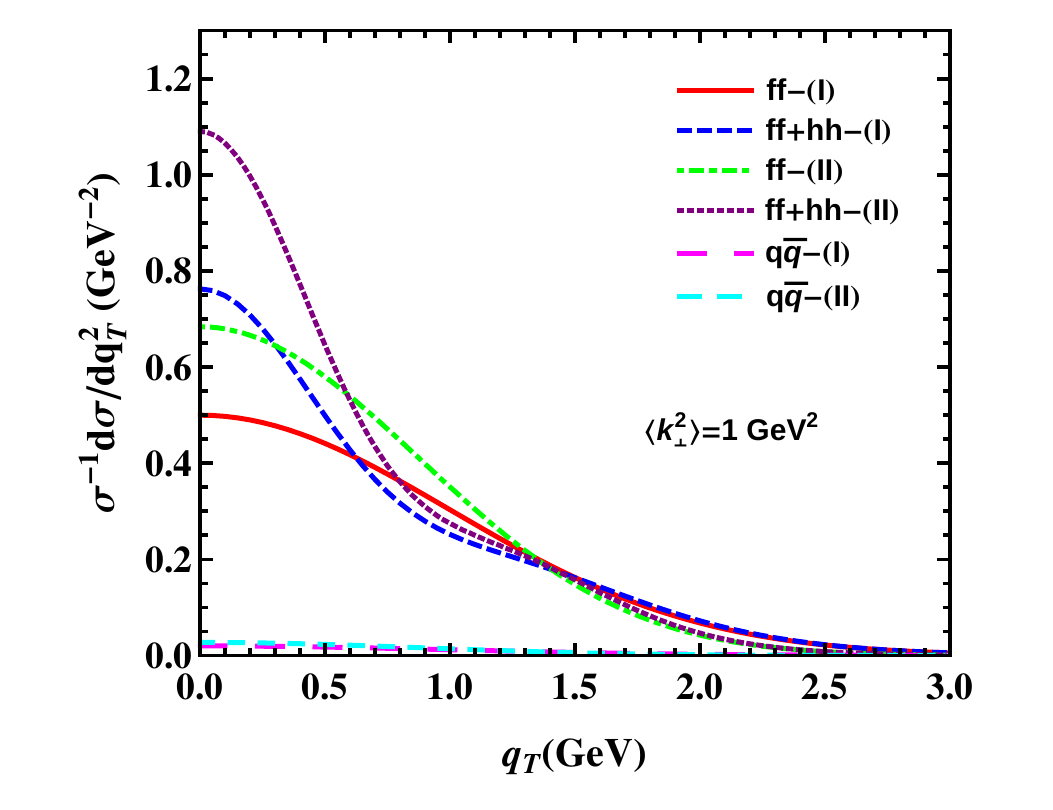}
\hspace{0.1cm}
\includegraphics[width=7.5cm,height=6cm,clip]{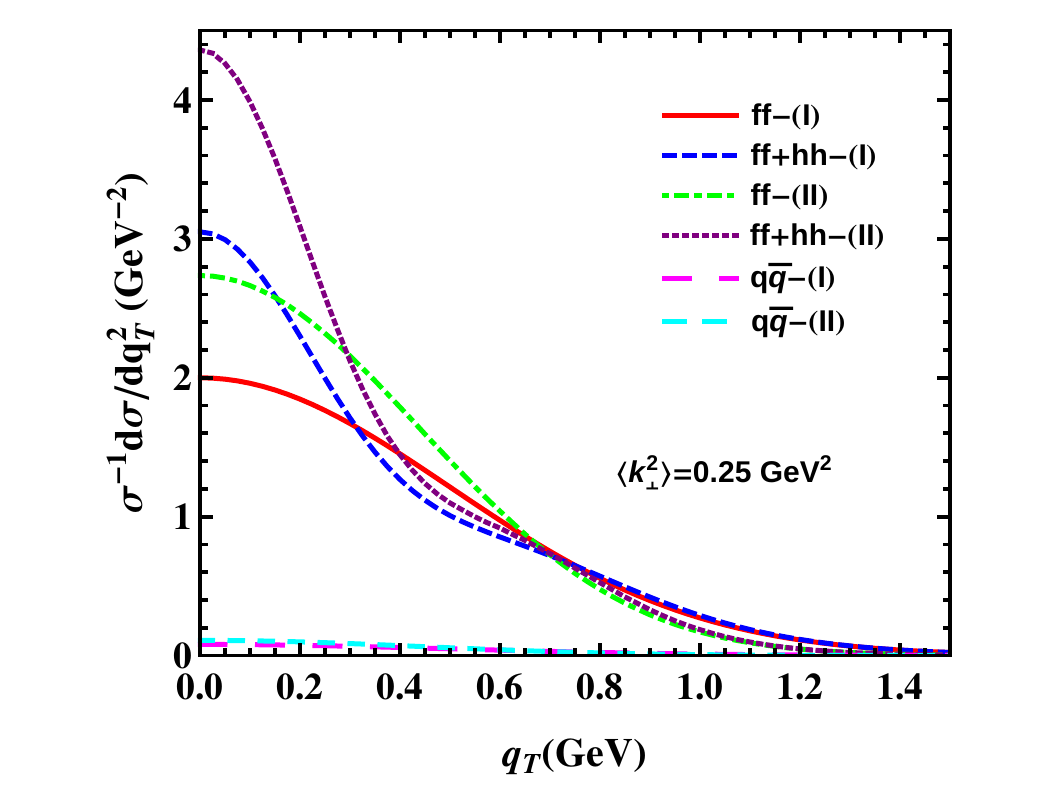}
\end{minipage}
\caption{\label{fig3}(color online) Same as in Fig. \ref{fig2} but for
$r=\frac{1}{3}$.}
\end{figure}

 The transverse momentum and rapidity distributions have been estimated in Model-I from 
Eq.\eqref{m1eq1} $\&$ 
 \eqref{m1eq2} and Model-II from Eq.\eqref{m2eq1} $\&$ \eqref{m2eq2}  by employing  a Gaussian 
 model. In Figs. \ref{fig2} and \ref{fig3} we have divided the result by the total cross section for the
kinematics of each experiment, as a result, we got overlapping curves
independent of the center-of-mass energy of the experiment and the mass of
the quarkonium. For different values of 
Gaussian width $\langle k^2_{\perp}\rangle$ and parameter
 $r$ the $q_T$  distribution of $J/\psi$ and  $\Upsilon$ has been shown in Fig. \ref{fig2} and 
\ref{fig3}. Fig. \ref{fig2} is for $r=2/3$ and Fig. \ref{fig3} is for
$r=1/3$. The invariant mass square of the heavy quark pair ($M^2$) which
 has very narrow range  from $4m^2_Q$ to $4m^2_{Q\bar{q}}$ in this model is used as a scale to 
evolve the pdfs for all experiments using DGLAP evolution equation in these
plots. We have noticed that linearly polarized gluons does not contribute to integrated $q_T$ 
cross section of  charmonium and bottomonium in DGLAP approach. The transverse momentum 
distribution of $J/\psi$ and  $\Upsilon$ in Model-I  is shown in Fig. \ref{fig2} and \ref{fig3} are  in agreement with  results \cite{prd86} for
 $\chi_{c,b0}$ quarkonium production obtained by NRQCD  framework. It is
 seen in the Fig. \ref{fig2} and \ref{fig3} that the inclusion of linearly polarized 
gluon contribution to the unpolarized
 gluon cross section have greatly modulated the transverse momentum distribution of 
charmonium and bottomonium
 mostly at low $q_T$, $q_T<0.5$~ GeV. So measuring the cross section of charmonium
production at low transverse momentum can help to disentangle the linearly
polarized gluon contribution. The transverse momentum dependent cross section of quarkonium is
higher in Model-II compared to Model-I. In Fig. \ref{fig3}, we have also
shown the $q\bar{q}$ contribution in quarkonium production. This is 
extremely small, compared to the gluon channel in the kinematics of the
experiments considered. Therefore we have not considered contribution of this
channel in other plots.

\begin{figure}[H]
\begin{minipage}[c]{0.99\textwidth}
\tiny{(a)}\includegraphics[width=7.5cm,height=6cm,clip]{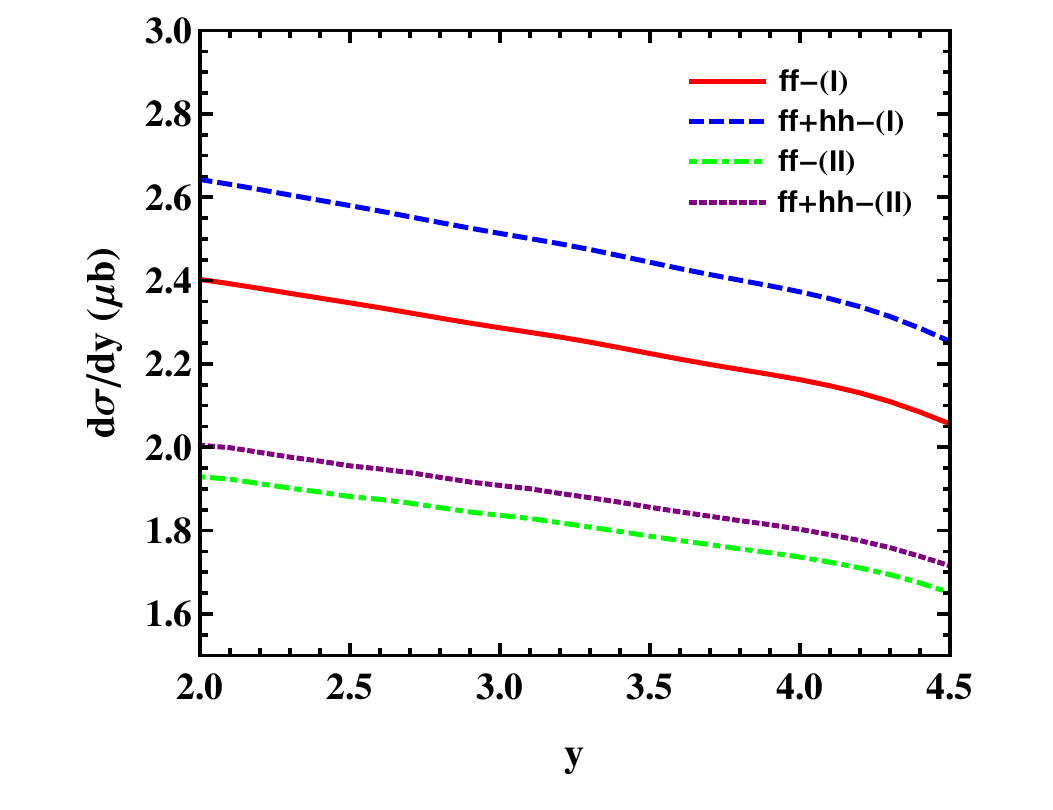}
\hspace{0.1cm}
\tiny{(b)}\includegraphics[width=7.5cm,height=6cm,clip]{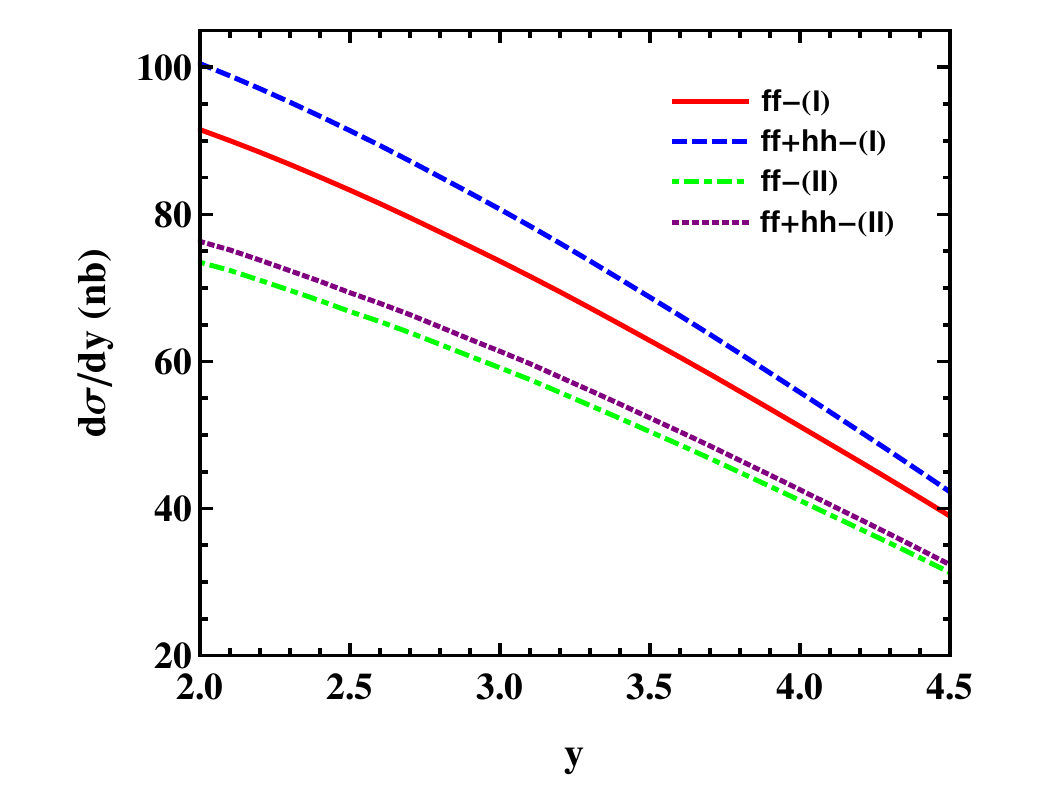}
\end{minipage}
\caption{\label{fig4}(color online) Rapidity ($y$) distribution of 
(a) $J/\psi$ (left panel) and  (b) $\Upsilon$ (right panel) in $\text{pp}\rightarrow Q\overline{Q}
+X$
at LHCb ($\sqrt{s}=7$ TeV) energy and $q_T$  integration range is from 0 to 0.5 GeV 
using DGLAP evolution approach. The convention in the figure for
 line styles is same as Fig. \ref{fig2}.}
\end{figure}

\begin{figure}[H]
\begin{minipage}[c]{0.99\textwidth}
\tiny{(a)}\includegraphics[width=7.5cm,height=6cm,clip]{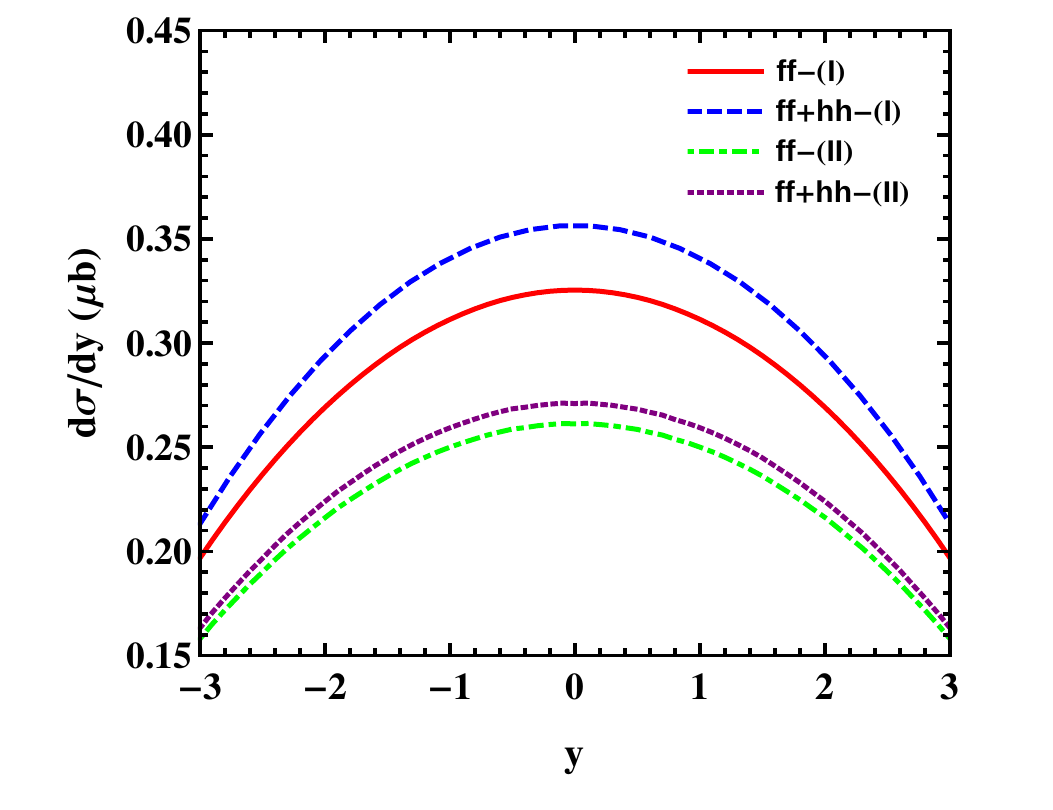}
\hspace{0.1cm}
\tiny{(b)}\includegraphics[width=7.5cm,height=6cm,clip]{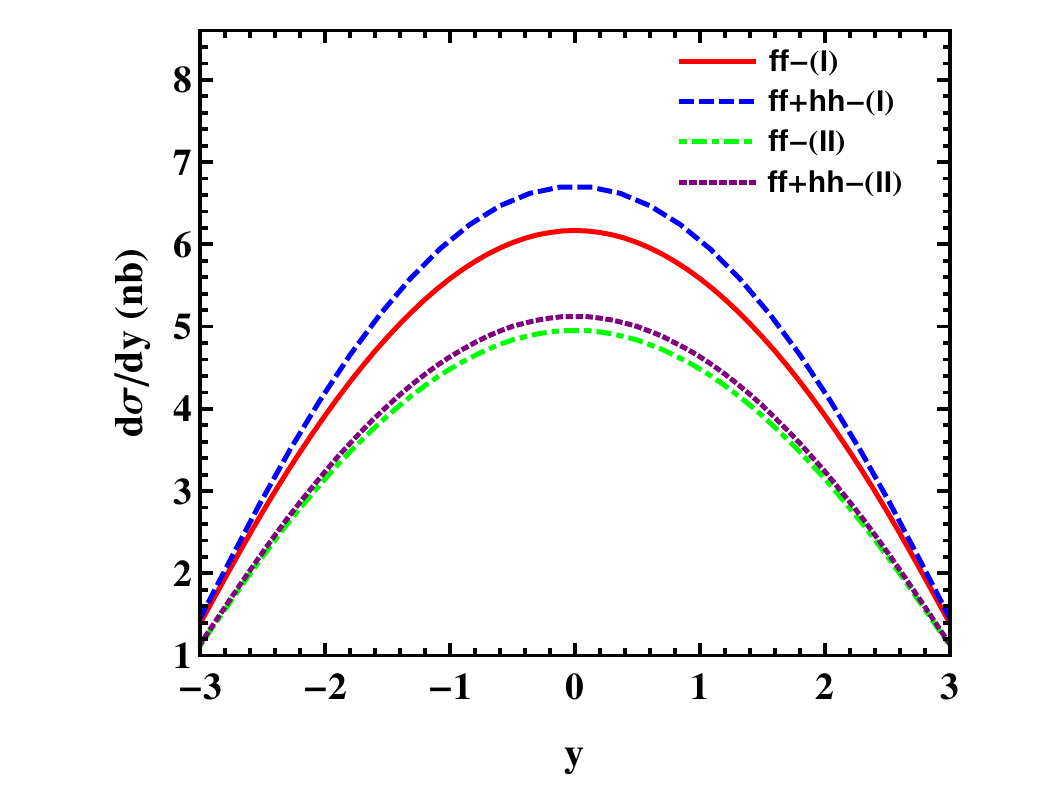}
\end{minipage}
\caption{\label{fig5}(color online) Rapidity ($y$) distribution of 
(a) $J/\psi$ (left panel) and  (b) $\Upsilon$ (right panel) in $\text{pp}\rightarrow Q\overline{Q}+X$
at RHIC ($\sqrt{s}=500$ GeV) energy and $q_T$  integration range is from 0 to 0.5 GeV using DGLAP evolution
approach. The convention in the figure for  line styles is same as Fig. \ref{fig2}.}
\end{figure}

\begin{figure}[H]
\begin{minipage}[c]{0.99\textwidth}
\tiny{(a)}\includegraphics[width=7.5cm,height=6cm,clip]{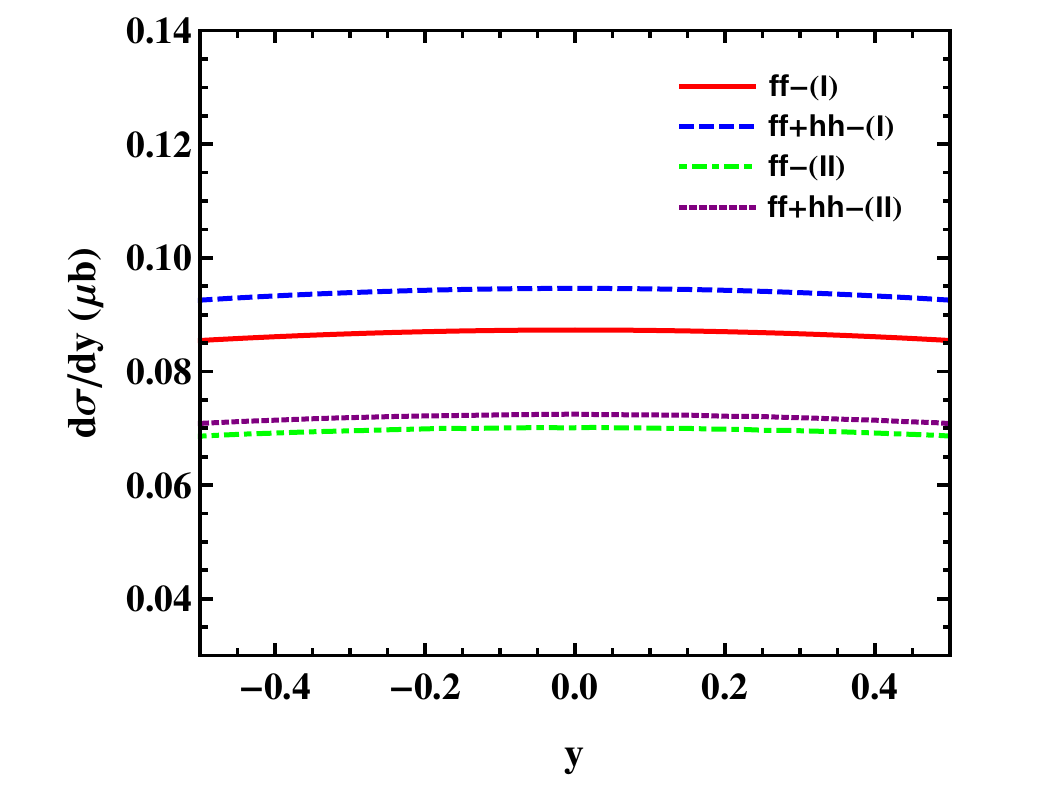}
\hspace{0.1cm}
\tiny{(b)}\includegraphics[width=7.5cm,height=6cm,clip]{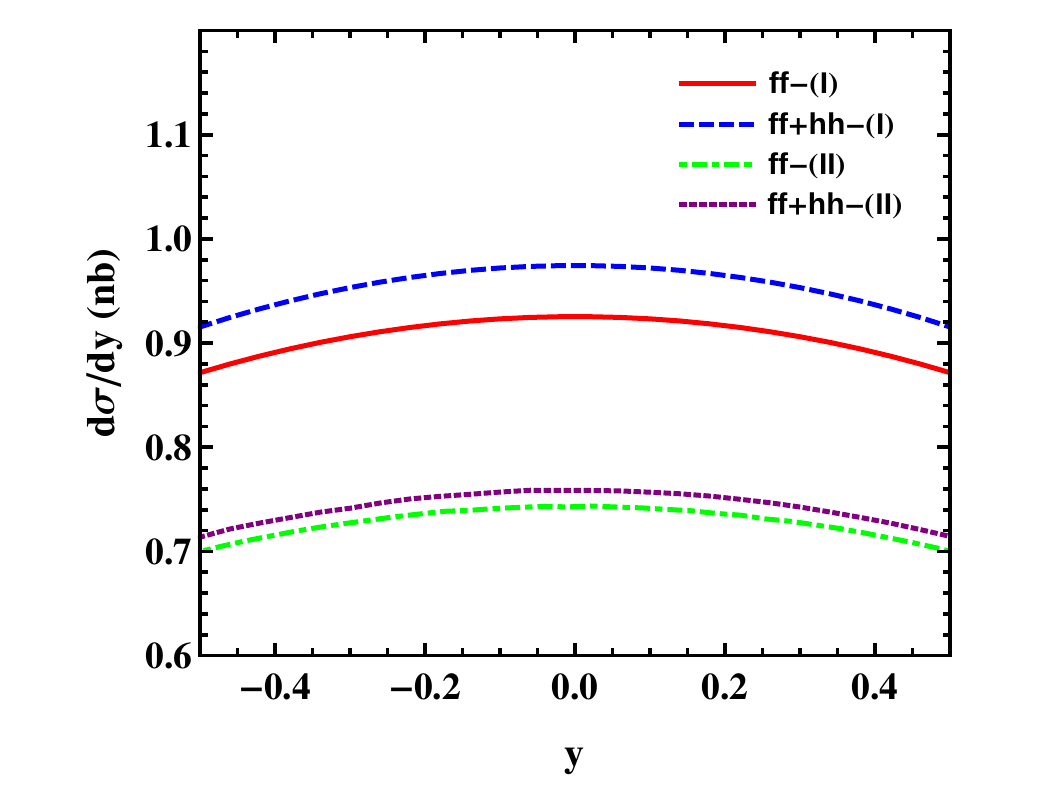}
\end{minipage}
\caption{\label{fig6}(color online). Rapidity ($y$) distribution of 
(a) $J/\psi$ (left panel) and  (b) $\Upsilon$ (right panel) in $\text{pp}\rightarrow Q\overline{Q}+X$
at AFTER ($\sqrt{s}=115$ GeV) energy and $q_T$  integration range is from 0 to 0.5 GeV using DGLAP evolution
approach. The convention in the figure for  line styles is same as Fig. \ref{fig2}.}
\end{figure}

Rapidity distribution of charmonium and bottomonium has been obtained in Model-I and II 
using DGLAP evolution which is shown in figures \ref{fig4}$-$\ref{fig6}.
Fig. \ref{fig4} is for the kinematics of LHCb, Fig. \ref{fig5} is for RHIC
  and Fig. \ref{fig6} is for the
kinematics of AFTER. We have chosen different rapidity range for different experiments, that are
given above. In order to show the effect of  linearly polarized gluons, we have chosen a
small $q_T$ bin, namely $0<q_T<0.5$ GeV in all these plots.  The cross section decreases 
with increasing rapidity. Rapidity distribution
is enhanced by considering the linearly polarized gluons apart from unpolarized gluons in 
the cross section. Moreover, the enhancement is more in
Model-I  compared to Model-II.  Furthermore, we have noticed that the 
$y$ distribution is independent of parameter $r$ and Gaussian width  $\langle k^2_{\perp}\rangle$.

\begin{figure}[H]
\begin{minipage}[c]{0.99\textwidth}
\tiny{(a)}\includegraphics[width=7.5cm,height=6cm,clip]{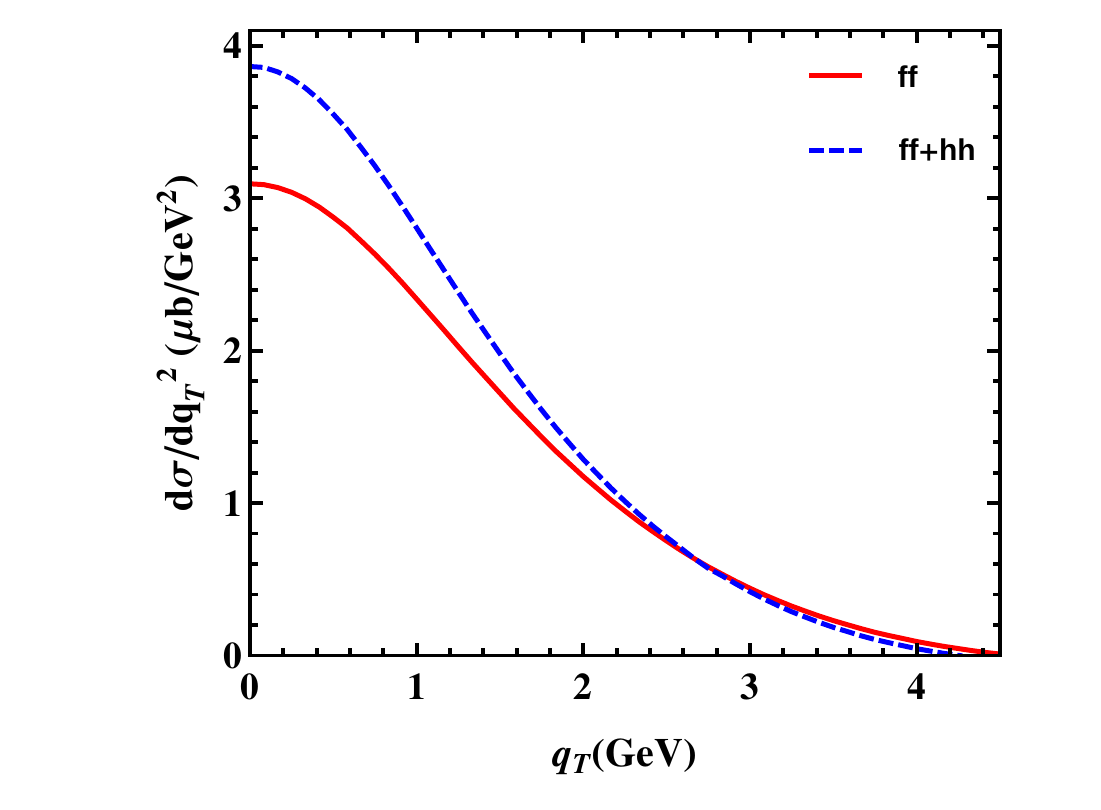}
\hspace{0.1cm}
\tiny{(b)}\includegraphics[width=7.5cm,height=6cm,clip]{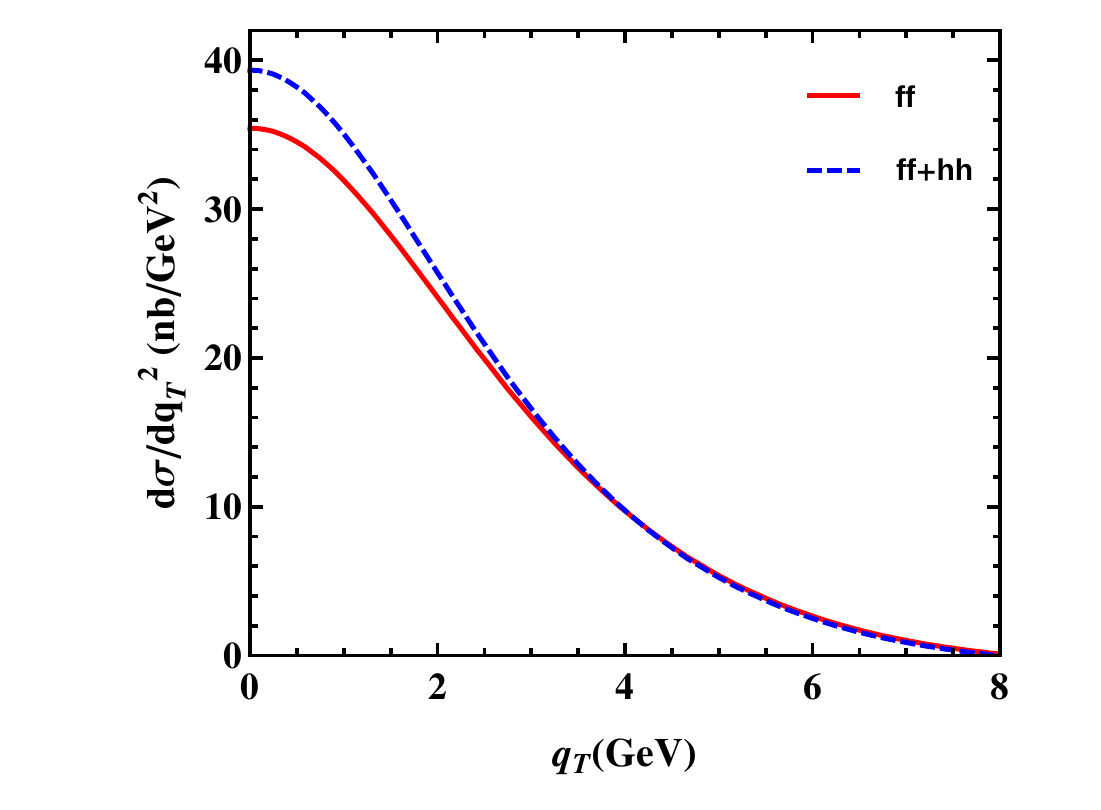}
\end{minipage}
\caption{\label{fig7}(color online). Differential cross section of  (a) $J/\psi$ (left panel)
and  (b) $\Upsilon$ (right panel) as function of $q_T$ 
in $\text{pp}\rightarrow Q\overline{Q}+X$ at LHCb ($\sqrt{s}=7$ TeV) energy 
using TMD evolution approach. The solid (ff) and dashed (ff+hh)  lines are obtained
by considering  unpolarized gluons only and unpolarized plus linearly polarized gluons 
respectively.}
\end{figure}

\begin{figure}[H]
\begin{minipage}[c]{0.99\textwidth}
\tiny{(a)}\includegraphics[width=7.5cm,height=6cm,clip]{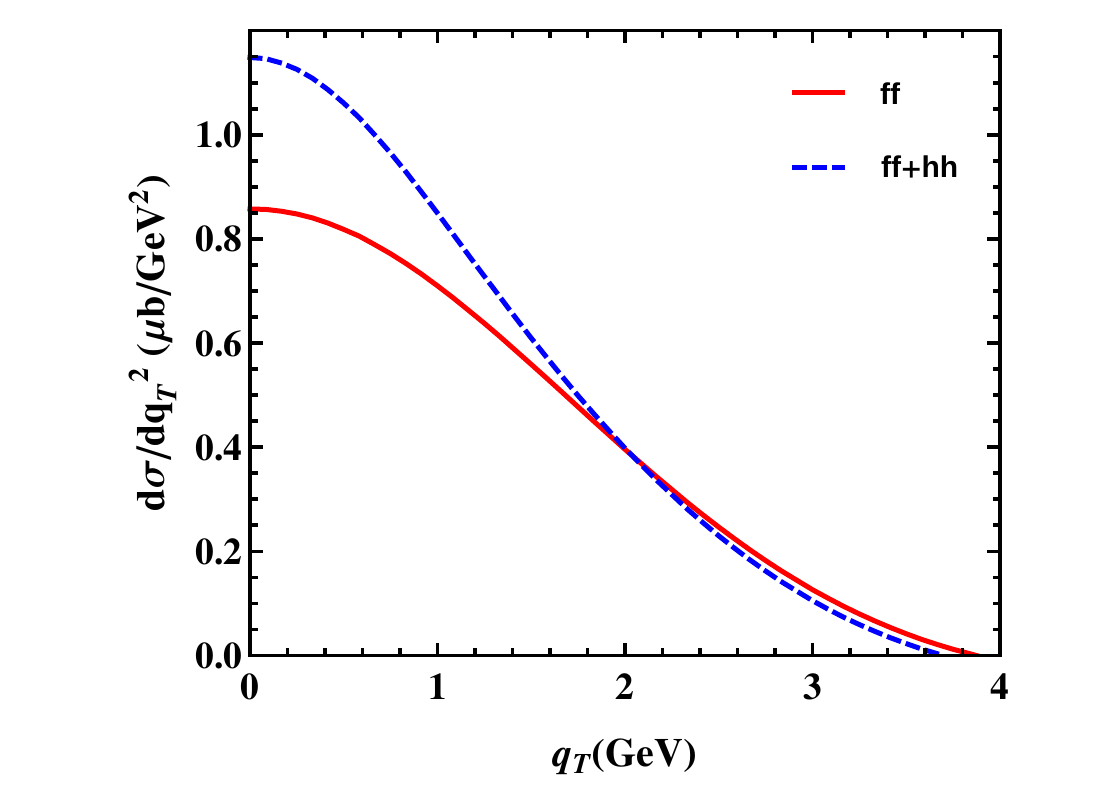}
\hspace{0.1cm}
\tiny{(b)}\includegraphics[width=7.5cm,height=6cm,clip]{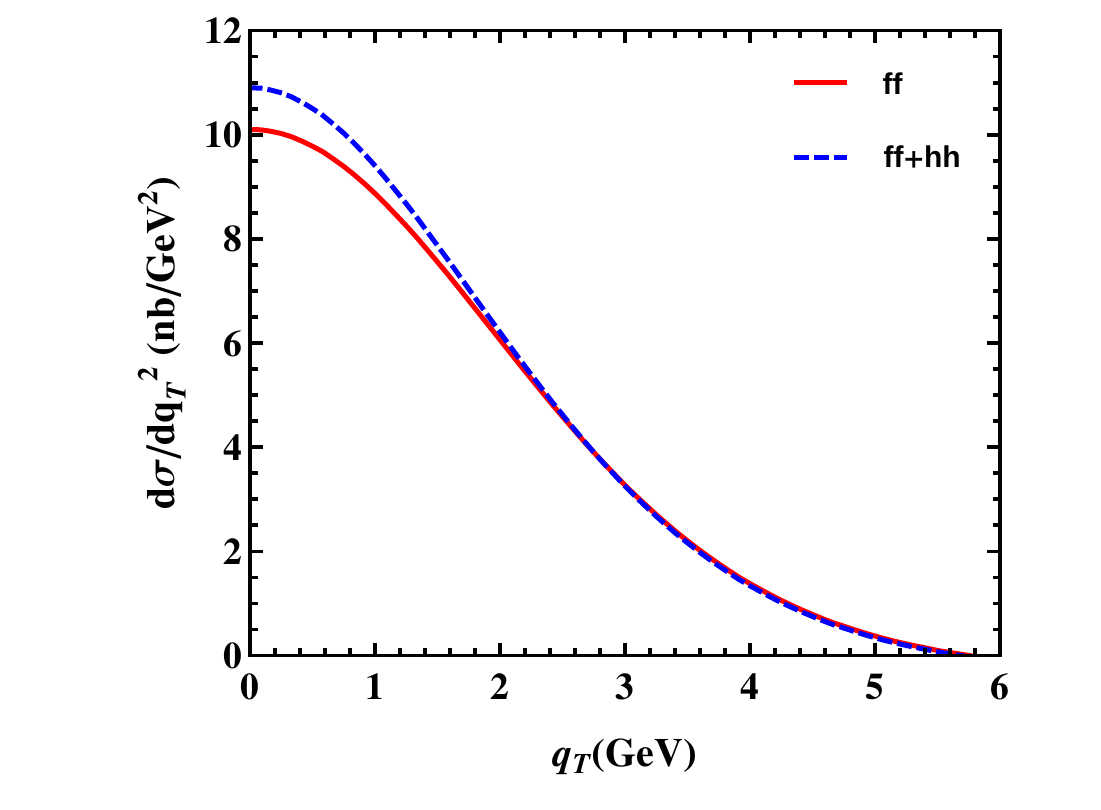}
\end{minipage}
\caption{\label{fig8}(color online).  Differential cross section of  (a) $J/\psi$ (left panel)
and  (b) $\Upsilon$ (right panel) as function of $q_T$
in $\text{pp}\rightarrow Q\overline{Q}+X$ at RHIC ($\sqrt{s}=500$ GeV) energy 
using TMD evolution approach.  The convention in the figure for  line styles is
same as Fig. \ref{fig7}.}
\end{figure}

\begin{figure}[H]
\begin{minipage}[c]{0.99\textwidth}
\tiny{(a)}\includegraphics[width=7.5cm,height=6cm,clip]{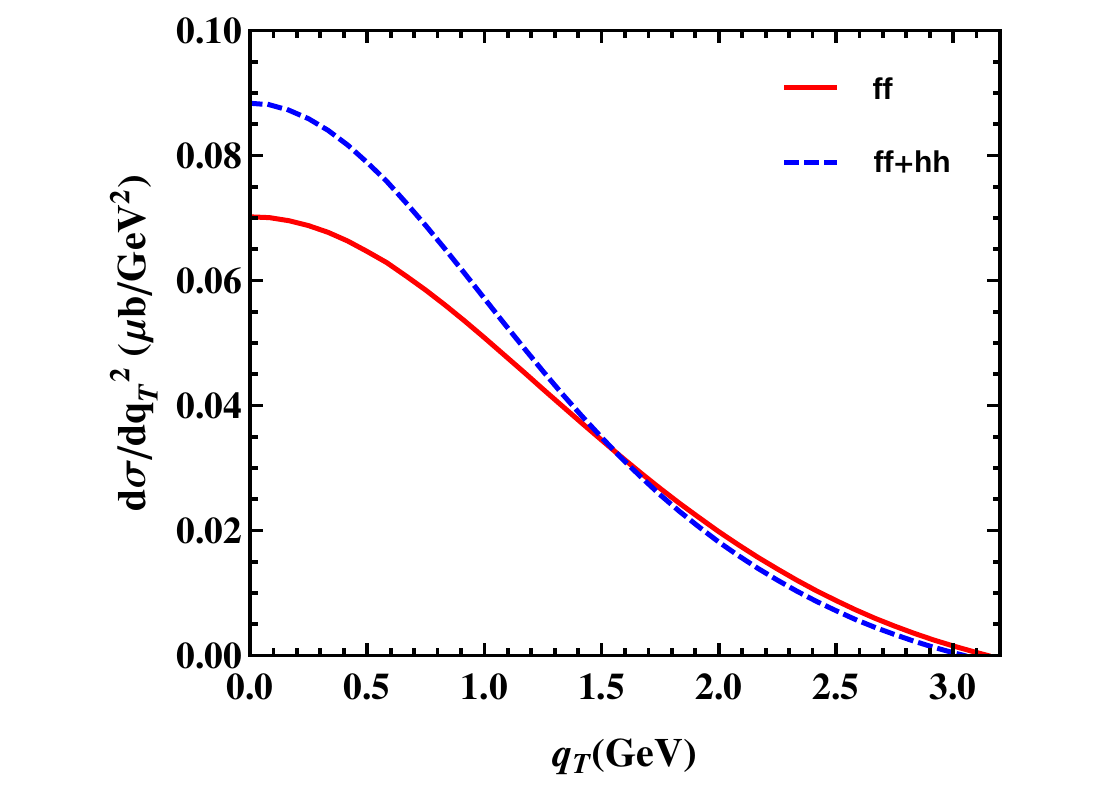}
\hspace{0.1cm}
\tiny{(b)}\includegraphics[width=7.5cm,height=6cm,clip]{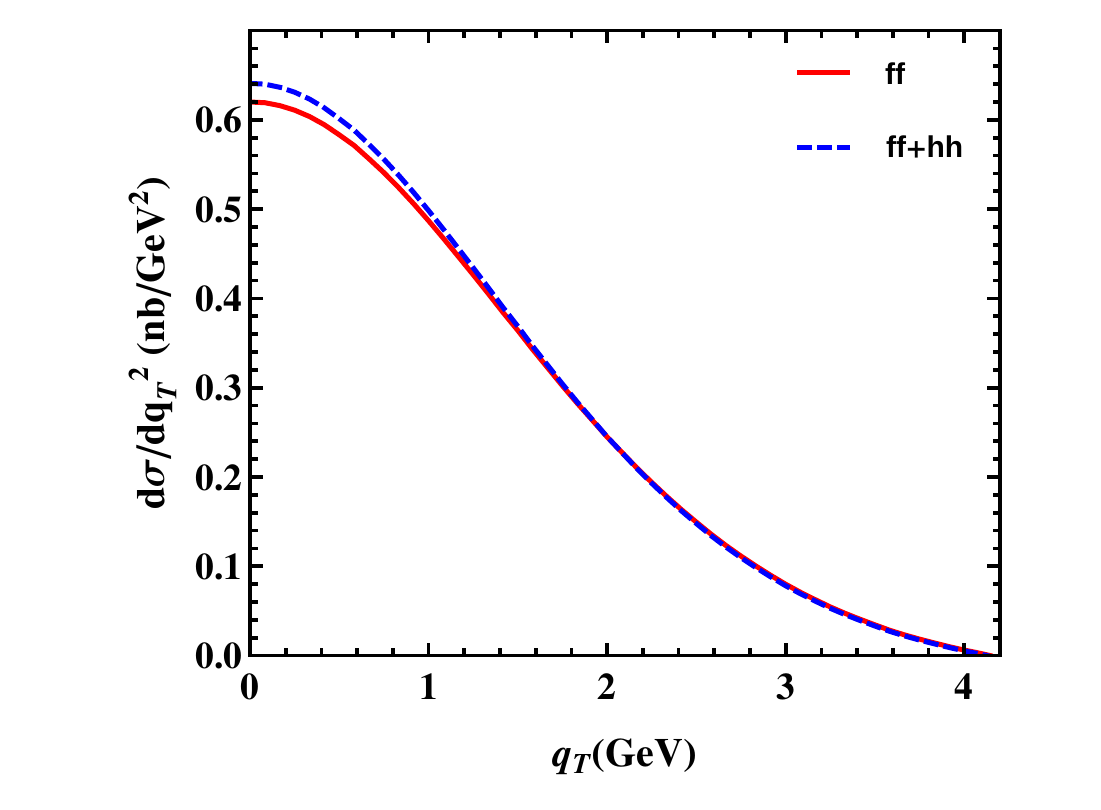}
\end{minipage}
\caption{\label{fig9}(color online). Differential cross section of  (a) $J/\psi$ (left panel)
and  (b) $\Upsilon$ (right panel) as function of $q_T$ 
in $\text{pp}\rightarrow Q\overline{Q}+X$ at AFTER ($\sqrt{s}=115$ GeV) energy 
using TMD evolution approach.  The convention in the figure for  line styles is
same as Fig. \ref{fig7}.}
\end{figure}

\begin{figure}[H]
\begin{minipage}[c]{0.99\textwidth}
\tiny{(a)}\includegraphics[width=7.5cm,height=6cm,clip]{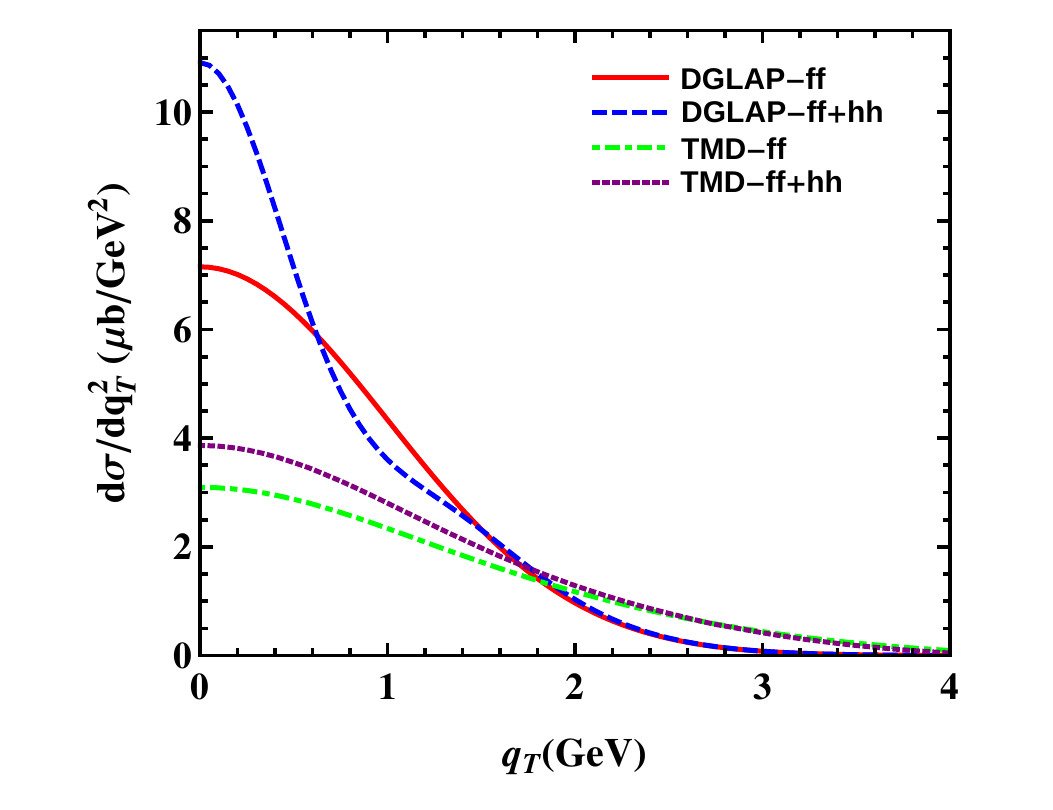}
\hspace{0.1cm}
\tiny{(b)}\includegraphics[width=7.5cm,height=6cm,clip]{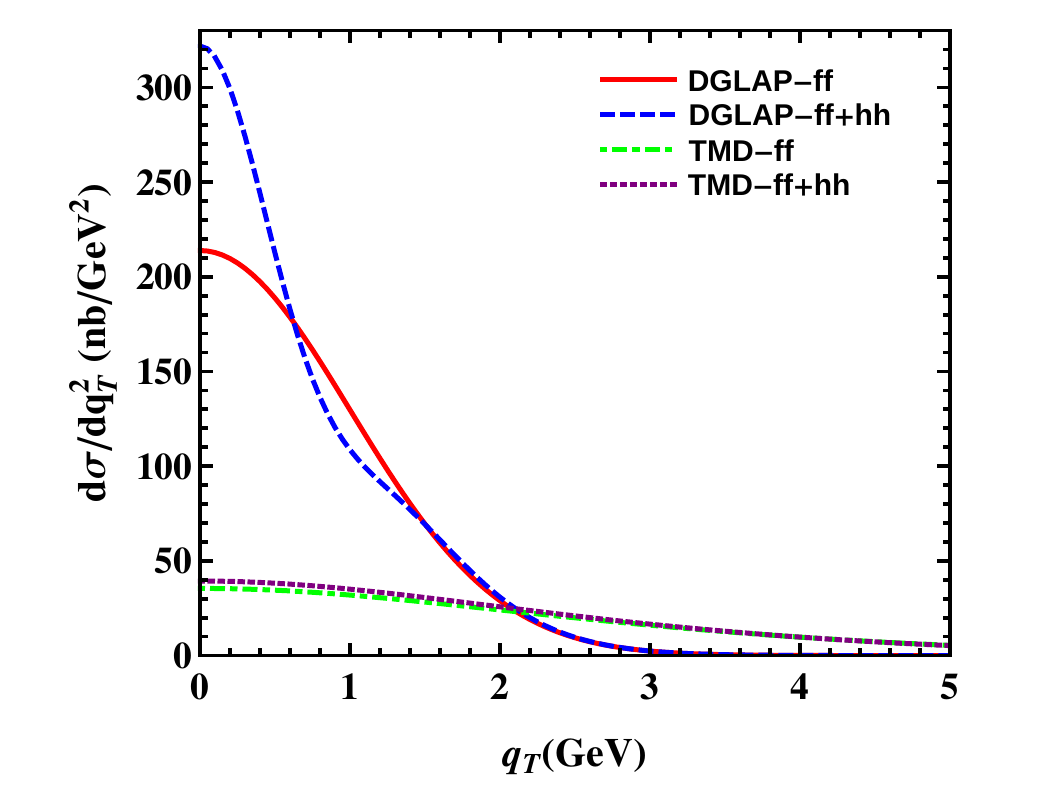}
\end{minipage}
\caption{\label{fig10}(color online).  Differential cross section of (a)  $J/\psi$ (left panel)
and (b)$\Upsilon$ (right panel) in $\text{pp}\rightarrow Q\overline{Q}+X$  at LHCb ($\sqrt{s}=7$ TeV).
The solid (DGLAP-ff) and dot dashed (TMD-ff) lines are obtained by considering
unpolarized gluons and quarks in DGLAP and only unpolarized gluons in TMD evolution  respectively.
The dashed (DGLAP-ff+hh) and tiny dashed (TMD-ff+hh) lines are obtained by  taking 
into account unpolarized gluons and quarks plus linearly polarized gluons in DGLAP and unpolarized 
plus linearly polarized gluons in TMD evolution respectively. We have chosen $r=\frac{1}{3}$
 and $\langle k^2_{\perp}\rangle=1~\mathrm{GeV}^2$ are taken in Model-I for DGLAP evolution.}
\end{figure}

Within TMD evolution formalism using Eq.\eqref{tmdevo}, the transverse momentum dependent 
cross section of quarkonium is shown in figures \ref{fig7}$-$\ref{fig9}.
Fig. \ref{fig7} is for the kinematics of LHCb, Fig. \ref{fig8} is for RHIC
and Fig. \ref{fig9} is for AFTER. In theses plots, we have integrated over
the rapidity in the ranges given above for different experiments. 
The effect of $h_1^{\perp g}$  in the  $q_T$ distribution in the TMD
evolution approach is not as dominant as in the DGLAP evolution approach,
particularly for $\Upsilon$, although it is sizable at low $q_T$. The mass of 
the $\Upsilon$ is more than that of $J/\psi$, and this effect is
suppressed by the mass. Only one loop in 
$\alpha_s$ has been taken to integrate the Sudakov factor in Eq.\eqref{sudakov}. 
The transverse momentum distribution of quarkonium is reduced in TMD evolution formalism compared 
to DGLAP evolution formalism. 
Fig. \ref{fig10} represents the decline of  $q_T$ distribution  in TMD approach
in 
contrast to DGLAP. Here we have chosen the kinematics of LHCb experiment.
The rapidity is integrated over the region  $y\in[2.0,4.5]$ 
TMD pdfs are evolved in TMD evolution from initial scale ($c/b_\ast$) to final scale ($Q$), 
where $Q$ has been set equal to quarkonium mass i.e., $Q=M$ which is the relevant scale 
for production of charmonium and bottomonium.

 
\section{Conclusion}\label{sec5}

Summarizing, we studied transverse momentum and rapidity distributions of  $J/\psi$ and $\Upsilon$
 in unpolarized proton-proton collision  within the formalism of transverse momentum dependent 
factorization. Since a long time, a lot of efforts have been put forward to understand the 
hadronization of  heavy quarks into mesons, both experimentally and theoretically. However, 
none of the models (CSM, NRQCD and CEM) could describe the transverse momentum dependent cross 
section of $J/\psi$  completely by fitting with experimental data \cite{lhcb}. Therefore, 
it would be interesting to include  the linearly polarized gluon contribution in the cross 
section to fit the experimental data to the extent of  reasonable accuracy.
On the other hand, by combining data from different experiments in different
kinematical regions one can quantify the magnitude of $h^{\perp g}_1$. Most
experiments measure spin and azimuthal asymmetries to probe the transverse
momentum dependent parton distributions and get information on the spin and
angular momentum structure of the hadrons. It is
also important to get a quantitative understanding of the unpolarized gluon TMD
$f^g_1(x,k_\perp)$. This lies at the denominator of the spin asymmetries and
its contribution is important in small $x$ region that is expected to play
an important role in collider experiments for example, in the future eRHIC. 
We employed color evaporation model (CEM), for its simplicity, to calculate
the differential cross section of quarkonium production and to illustrate the effect
of linearly polarized gluons distribution. We
studied the effect of TMD evolution at LHCb, RHIC and AFTER energies. We
found that contribution from the $q{\bar q}$ channel is very small compared
to the gluon channel.  We observed that 
the inclusion of linearly polarized gluons significantly modulated the transverse momentum 
distribution of  $J/\psi$ and $\Upsilon$ at low $q_T$. We further noticed that rapidity 
distribution has been enhanced by taking the presence of linearly polarized gluons into account
inside the unpolarized proton. By this we conclude that charmonium and bottomonium production
via gluon fusion process is a very useful tool to probe the unpolarized gluon TMD and linearly 
polarized gluons distribution.

\section*{ACKNOWLEDGEMENTS}
S.Rajesh acknowledges Sreeraj Nair for his help in Fortran code. Vikash Kumar Ojha is thanked
 for his help rendered during initial stage of this project.

\end{document}